\documentclass[12pt,journal,draftcls,a4paper,onecolumn]{IEEEtran}
\usepackage{algorithm}
\usepackage{algorithmic}
\usepackage{graphicx}
\usepackage[cmex10]{amsmath}
\usepackage{amsfonts}
\usepackage{amssymb}
\usepackage{subfigure}
\usepackage{color}
\usepackage[noadjust]{cite}
\usepackage{verbatim}
\usepackage{dsfont}
\usepackage{epic,eepic,pstricks}

\graphicspath{{figures/}}

\title{Estimating the Granularity Coefficient \\of a Potts-Markov Random Field \\within an MCMC Algorithm}
\author{Marcelo Pereyra, Nicolas Dobigeon, \\Hadj Batatia and
Jean-Yves Tourneret
\\
\normalsize University of Toulouse, IRIT/INP-ENSEEIHT/T\'eSA\\
2 rue Camichel, 31071 Toulouse, France. \\
\small\texttt{\{Marcelo.Pereyra,Nicolas.Dobigeon,\\Hadj.Batatia,Jean-Yves.Tourneret\}@enseeiht.fr}}

\begin{document}

\maketitle
\begin{abstract}
This paper addresses the problem of estimating the Potts parameter $\beta$ jointly with the unknown parameters of a Bayesian model within a Markov chain Monte Carlo (MCMC) algorithm. Standard MCMC methods cannot be applied to this problem because performing inference on $\beta$ requires computing the intractable normalizing constant of the Potts model. In the proposed MCMC method the estimation of $\beta$ is conducted using a likelihood-free Metropolis-Hastings algorithm. Experimental results obtained for synthetic data show that estimating $\beta$ jointly with the other unknown parameters leads to estimation results that are as good as those obtained with the actual value of $\beta$. On the other hand, assuming that the value of $\beta$ is known can degrade estimation performance significantly if this value is incorrect. To illustrate the interest of this method, the proposed algorithm is successfully applied to real bidimensional SAR and tridimensional ultrasound images.

\end{abstract}
\section{Introduction}
\label{sec:intro} Modeling spatial correlation in images is
fundamental in many image processing applications. Markov random
fields (MRF) have been recognized as efficient tools for capturing
these spatial correlations
\cite{Geman1984,LiBook,Cordero2012,Mahapatra2012,Katsuki2012,Jain2012}.
One particular MRF often used for Bayesian classification and
segmentation is the Potts model, which generalizes the binary Ising
model to arbitrary discrete vectors. The amount of spatial
correlation introduced by this model is controlled by the so-called
\textit{granularity coefficient} $\beta$. In most applications this
important parameter is set heuristically by cross-validation.

This paper studies the problem of estimating the Potts coefficient $\beta$ jointly with the other unknown parameters of a standard Bayesian image classification or segmentation problem. More precisely, we consider Bayesian models defined by a conditional observation model with unknown parameters and a discrete hidden label vector $\boldsymbol{z}$ whose prior distribution is a Potts model with hyperparameter $\beta$ (this Bayesian model is defined in Section \ref{sec:Bayesian Modeling}). From a methodological perspective, inference on $\beta$ is challenging because the distribution $f(\boldsymbol{z},\beta)$ depends on the normalizing constant of the Potts model (hereafter denoted as $C(\beta)$), which is generally intractable. This problem has received some attention in the recent image processing literature, as it would lead to fully unsupervised algorithms \cite{Risser2009,Risser2010,McGrory2009,Pelizzari2010,Picco2011}.

In this work we focus on the estimation of $\beta$ within a Markov chain Monte Carlo (MCMC) algorithm that handles $2$D or $3$D data sets \cite{Mignotte2007,Mignotte2010,Kayabol2009,Eches2010tgrs,PereyraTMIC2011}. MCMC methods are powerful tools to handle Bayesian inference problems for which the minimum mean square error (MMSE) or the maximum a posteriori (MAP) estimators are difficult to derive analytically. MCMC methods generate samples that are asymptotically distributed according to the joint posterior of the unknown parameters. These samples are then used to approximate the Bayesian estimators. However, standard MCMC methods cannot be applied directly to Bayesian problems based on the Potts model. Indeed, inference on $\beta$ requires computing the normalizing constant of the Potts model $C(\beta)$, which is generally intractable. Specific MCMC algorithms have been designed to estimate Markov field parameters in \cite{Descombes1999,Murray2004} and more recently in \cite{Risser2009,Risser2010}. A variational Bayes algorithm based on an approximation of $C(\beta)$ has also been recently proposed in  \cite{McGrory2009}. Maximum likelihood estimation of $\beta$ within expectation-maximization (EM) algorithms has been studied in \cite{Pelizzari2010,Picco2011,Yongfeng2005}. The strategies involved in these works for avoiding computing the normalizing constant $C(\beta)$ are summarized below.

\subsection{Pseudo-likelihood estimators}
One possibility to avoid the computation of $C(\beta)$ is to eliminate it from the posterior distribution of interest. More precisely, one can think of defining a prior distribution $f(\beta)$ such that the normalizing constant cancels out from the posterior (i.e., $f(\beta)  \propto C(\beta)\boldsymbol{1}_{\mathcal{R}^+}(\beta)$), resulting in the so-called \emph{pseudo-likelihood} estimators \cite{Besag1975}. Although analytically convenient this approach generally does not lead to a satisfactory posterior density and results in poor estimation \cite{Geyer1992}. Also, as noticed in \cite{Murray2004} such a prior distribution generally depends on the data since the normalizing constant $C(\beta)$ depends implicitly on the number of observations (priors that depend on the data are not recommended in the Bayesian paradigm \cite[p. 36]{Gelman_book}).

\subsection{Approximation of $C(\beta)$}
Another possibility is to approximate the normalizing constant $C(\beta)$. Existing approximations can be classified into three categories: based on analytical developments, on sampling strategies or on a combination of both. A survey of the state-of-the-art approximation methods up to 2004 has been presented in \cite{Murray2004}. The methods considered in \cite{Murray2004} are the mean field, the tree-structured mean field and the Bethe energy (loopy Metropolis) approximations, as well as two sampling strategies based on Langevin MCMC algorithms. More recently, exact recursive expressions have been proposed to compute $C(\beta)$ analytically \cite{McGrory2009}. However, to our knowledge, these recursive methods have only been successfully applied to small problems (i.e., for MRFs of size smaller than $40 \times 40$) with reduced spatial correlation $\beta < 0.5$.

Another sampling-based approximation consists in estimating $C(\beta)$ by Monte Carlo integration \cite[Chap. 3]{Robert}, at the expense of very substantial computation and possibly biased estimations (bias arises from the estimation error of $C(\beta)$). Better results can be obtained by using importance or path sampling methods \cite{Gelman1998}. These methods have been applied to the estimation of $\beta$ within an MCMC image processing algorithm in \cite{Descombes1999}. Although more precise than Monte Carlo integration, approximating $C(\beta)$ by importance or path sampling still requires substantial computation and is generally unfeasible for large fields. This has motivated recent works that reduce computation by combining importance sampling with analytical approximations. More precisely, approximation methods that combine importance sampling with extrapolation schemes have been proposed for the Ising model (i.e., a $2$-state Potts model) in \cite{Risser2009} and for the $3$-state Potts model in \cite{Risser2010}. However, we have found that this extrapolation technique introduces significant bias \cite{Pereyra_TIP_TechReport_2012}.

\subsection{Auxiliary variables and perfect sampling}
Recent works from computational statistics have established that it is possible to avoid computing $C(\beta)$ within a Metropolis-Hastings MCMC algorithm \cite{Robert} by introducing carefully selected auxiliary random variables \cite{Moller2006,Andrieu2010}. In the work of Moller \emph{et. al.} \cite{Moller2006}, an auxiliary vector $\boldsymbol{w}$ distributed according to the same distribution as the label vector $\boldsymbol{z}$ (i.e., $f(\boldsymbol{z}|\beta)$) is introduced. Metropolis-Hastings algorithms that do not require computing $C(\beta)$ are then proposed to sample the joint distribution $f(\beta, \boldsymbol{w}| \boldsymbol{z})$, which admits the exact desired posterior density $f(\beta|\boldsymbol{z})$ as marginal distribution \cite{Moller2006}. Unfortunately this method suffers from a very low acceptance ratio that degrades severely as the dimension of $\boldsymbol{z}$ increases, and is therefore unsuitable for image processing applications \cite{Pereyra_TIP_TechReport_2012}.
Novel auxiliary variable methods with considerably better acceptance ratios have been proposed in \cite{Andrieu2010} by
using several auxiliary vectors and sequential Monte Carlo samplers \cite{DelMoral2006}. These methods could be interesting for
estimating the Potts coefficient $\beta$. However they will not be considered in this work because they require substantial computation and are generally too costly for image processing applications.
An alternative auxiliary variable method based on a one-sample estimator of the ratio $\frac{C(\beta)}{C(\beta^*)}$ has been proposed in \cite{Murray2006} and recently been improved by using several auxiliary vectors and sequential Monte Carlo samplers in \cite{Everitt2012} (the ratio $\frac{C(\beta)}{C(\beta^*)}$ arises in the MCMC algorithm defined in Section \ref{ssec:beta}). More details on the application of \cite{Murray2006} to the estimation of the Potts coefficient $\beta$ are provided in a separate technical report \cite{Pereyra_TIP_TechReport_2012}.

\subsection{Likelihood-free methods}
Finally, it is possible to avoid computing the normalizing constant $C(\beta)$ by using likelihood-free MCMC methods \cite{Marjoram2003}. These methods substitute the evaluation of intractable likelihoods within a Metropolis-Hastings algorithm by a simulation-rejection scheme. More precisely, akin to the auxiliary variable method \cite{Moller2006}, an auxiliary vector $\boldsymbol{w}$ distributed according to the likelihood $f(\boldsymbol{z}|\beta)$ is introduced. Two-step Metropolis-Hastings algorithms that do not require evaluating $f(\boldsymbol{z}|\beta)$ (nor $C(\beta)$) can then be considered to generate samples that are asymptotically distributed according to the exact posterior distribution $f(\beta | \boldsymbol{z})$ \cite{Marjoram2003}. Although generally unfeasible\footnote{In spite of being theoretically correct, exact likelihood-free algorithms suffer from several major shortcomings that make them generally impractical (see Section \ref{subsec:auxiliary} for more details).}, these exact methods have given rise to the \emph{approximative Bayesian computation} (ABC) framework \cite{Marin2011}, which studies likelihood-free methods to generate samples from approximate posterior densities $f_\epsilon (\beta | \boldsymbol{z}) \approx f (\beta | \boldsymbol{z})$ at a reasonable computational cost. To our knowledge these promising techniques, that are increasingly regarded as ``\emph{the most satisfactory approach to intractable likelihood problems}'' \cite{Marin2011}, have not yet been applied to image processing problems.

The main contribution of this paper is to propose an ABC MCMC algorithm for the joint estimation of the label vector $\boldsymbol{z}$, the granularity coefficient $\beta$ and the other unknown parameters of a Bayesian model. The estimation of $\beta$ is included within an MCMC algorithm through an ABC method particularly adapted to the Potts model and to large data sets. It is shown that the estimation of $\beta$ can be easily integrated to existing MCMC algorithms where $\beta$ was previously assumed known. Applications to large $2$D and $3$D images illustrate the performance of the proposed method. T

The remainder of the paper is organized as follows: Bayesian models considered in this work are defined in Section II. Section III describes a generic hybrid Gibbs sampler to generate samples asymptotically distributed according to the approximate posterior distribution of these Bayesian models. The estimation of $\beta$ using a likelihood-free algorithm is discussed in detail in Section IV. Experiments on synthetic and real data are presented in Sections V and VI respectively. Conclusions are finally reported in Section VI.

\section{Bayesian Model} \label{sec:Bayesian Modeling}
\label{sec:probStatement} Let $r_n \in \mathbb{R}^{+}$ denote the $n$th
observation, or voxel, in a lexicographically vectorized image
$\boldsymbol{r}=(r_1,\ldots,r_N)^T \in \mathbb{R}^{N}$. We assume that $\boldsymbol{r}$ is made up by multiple regions,
characterized by their own statistics. More precisely, $\boldsymbol{r}$ is
assumed to be associated with $K$ stationary classes $\{\mathcal{C}_1, \ldots, \mathcal{C}_K\}$ such that the observations in the $k$th class are fully described by the following conditional observation model
\begin{equation} \label{mixture}
r_n|z_n = k \sim f\left(r_n\boldsymbol{|\theta}_k\right)
\end{equation}
where $f\left(r_n\boldsymbol{|\theta}_k\right)$ denotes a generic
observation model with parameter vector $\boldsymbol{\theta}_k$ characterizing the class $\mathcal{C}_k$. Finally, a label vector $\boldsymbol{z}=\left(z_1,\ldots,z_N\right)^T$ is introduced to map observations $\boldsymbol{r}$ to classes $\mathcal{C}_1, \ldots, \mathcal{C}_K$ (i.e., $z_n = k$ if and only if $r_n \in \mathcal{C}_k$).

Several works have established that a Potts model can be used to enhance the fact that the probability $\mathrm{P}[z_n = k]$ of a given voxel is related to the probabilities of its neighbors. As explained previously, the amount of spatial correlation between adjacent image pixels introduced by the Potts model is controlled by the granularity coefficient $\beta$. Existing image classification and segmentation methods have mainly studied the estimation of the class parameter vector $\boldsymbol{\theta} = (\boldsymbol{\theta}_1^T,\ldots,\boldsymbol{\theta}_K^T)^T$ and the label vector $\boldsymbol{z}$ conditionally to a known value of $\beta$. However, setting $\beta$ incorrectly can degrade the estimation of $\boldsymbol{\theta}$ and $\boldsymbol{z}$ significantly. Moreover, fixing the value of $\beta$ a priori is difficult because different images can have different spatial organizations. This paper considers the problem of estimating the unknown parameter vectors $\boldsymbol{\theta}$ and $\boldsymbol{z}$ jointly with $\beta$. This problem is formulated in a Bayesian framework which requires to define the likelihood of the observation vector $\boldsymbol{r}$ and the priors for the unknown parameters $\boldsymbol{\theta}$, $\boldsymbol{z}$ and $\beta$.

\subsection{Likelihood}\label{ssec:Likelihood}
Assuming that the observations $r_n$ are independent conditionally
to the label vector $\boldsymbol{z}$, the likelihood function
associated with the image $\boldsymbol{r}$ is
\begin{equation} \label{likelihood}
f(\boldsymbol{r}|\boldsymbol{\theta},\boldsymbol{z},\beta) = f(\boldsymbol{r}|\boldsymbol{\theta},\boldsymbol{z}) = \prod_{k=1}^K \,\,\prod_{\{n | z_n = k\}} f(r_n|\boldsymbol{\theta}_k)
\end{equation}
where $f(r_n|\boldsymbol{\theta}_k)$ is the generic probability density function associated with the observation
model introduced in \eqref{mixture}.

\subsection{Parameter priors}\label{ssec:Parameter priors}
\subsubsection{Labels}
\begin{figure}
  \centerline{\includegraphics[width=5.7cm]{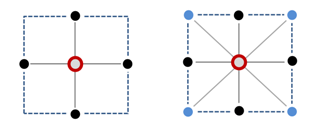}}
  \caption{4-pixel (left) and 8-pixel (right) neighborhood structures. The pixel considered appears as a void red circle whereas its neighbors are depicted in full black and blue.}\label{fig:neighborhood2D}
\end{figure}

\begin{figure}
  \centerline{\includegraphics[width=5.7cm]{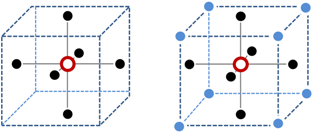}}
  \caption{6-voxel (left) and 14-voxel (right) neighborhood structures. The considered voxel appears as a void red circle whereas its neighbors are depicted in full black and blue.}\label{fig:neighborhood3D}
\end{figure}

It is natural to consider that there are some correlations between the
characteristics of a given voxel and those of its neighbors. Since
the seminal work of Geman \cite{Geman1984}, MRFs have become very
popular to introduce spatial correlation in images. MRFs assume that
the distribution of a pixel conditionally to all other pixels of the
image equals the distribution of this pixel conditionally to its
neighbors. Consequently, it is important to properly define the
neighborhood structure. The neighborhood relation between two pixels
(or voxels), $i$ and $j$, has to be symmetric: if $i$ is a neighbor of
$j$ then $j$ is also a neighbor of $i$. There are several
neighborhood structures that have been used in the literature. In
the bidimensional case, neighborhoods defined by the four or eight
nearest voxels represented in Fig. \ref{fig:neighborhood2D} are the
most commonly used. Similarly, in the tridimensional case the most
frequently used neighborhoods are defined by the six or fourteen
nearest voxels represented in Fig \ref{fig:neighborhood3D}. In the
rest of this paper $4$-pixel and $6$-voxel neighborhoods will be considered for
$2$D and $3$D images, respectively. Therefore,
the associated set of neighbors, or cliques, have vertical,
horizontal and depth configurations (see
\cite{Geman1984} for more details).

Once the neighborhood structure has been established, the MRF can be
defined. Let $z_n$ denote the random variable indicating the class
of the $n$th image voxel. The whole set of random variables
$z_1,z_2,\ldots,z_N$ forms a random field. An MRF is obtained
when the conditional distribution of $z_n$ given the other pixels
$\boldsymbol{z}_{-n} =(z_1,\ldots,z_{n-1},z_{n+1},\ldots,z_N)$ only depends on its neighbors
$\boldsymbol{z}_{\mathcal{V}(n)}$, i.e.,
\begin{equation}
f\left(z_n|\boldsymbol{z}_{-n}\right) =
f\left(z_n|\boldsymbol{z}_{\mathcal{V}(n)}\right)
\end{equation}
where $\mathcal{V}(n)$ is the index set of the neighbors of the
$n$th voxel, $\boldsymbol{z}_{-n}$ denotes the vector $\boldsymbol{z}$ whose $n$th element has been removed
and $\boldsymbol{z}_{\mathcal{V}(n)}$ is the sub-vector of
$\boldsymbol{z}$ composed of the elements whose indexes belong to
$\mathcal{V}(n)$.

In the case of $K$ classes, the random variables
$z_1,z_2,\ldots,z_N$ take their values in the finite set
$\{1,\ldots,K\}$. The resulting MRF (with discrete values) is a
Potts-Markov field, which generalizes the binary Ising model to arbitrary discrete vectors. In this study $2$D and $3$D Potts-Markov
fields will be considered as prior distributions for
$\boldsymbol{z}$. More precisely, $2$D MRFs are considered for
single-slice ($2$D) images whereas $3$D MRFs are investigated for
multiple-slice ($3$D) images. Note that Potts-Markov fields are
particularly well suited for label-based segmentation as explained
in \cite{Wu1982}. By the Hammersley-Clifford theorem the
corresponding prior for $\boldsymbol{z}$ can be expressed as
follows
\begin{equation} \label{Potts}
f(\boldsymbol{z}|\beta) = \frac{1}{C(\beta)} \exp{\left[\Phi_{\beta}(\boldsymbol{z})\right]}
\end{equation}
where
\begin{equation} \label{eq:potential}
    \Phi_{\beta}(\boldsymbol{z}) = \sum_{n=1}^N\sum_{n'\in \mathcal{V}(n)} \beta \delta (z_{n} - z_{n'})
\end{equation}
and where $\delta(\cdot)$ is the Kronecker function, $\beta$ is the
granularity coefficient and $C(\beta)$ is the normalizing constant
or partition function \cite{Kindermann1980}
\begin{equation}
  \label{eq:partition}
  C(\beta) = \sum_{\boldsymbol{z} \in \left\{1,\ldots,K\right\}^n}
  \exp\left[\Phi_{\beta}\left(\boldsymbol{z}\right)\right].
\end{equation}
As explained previously, the normalizing constant $C(\beta)$ is generally intractable even for $K=2$ because the number of summands in \eqref{eq:partition} grows exponentially with the size of $\boldsymbol{z}$ \cite{Vincent2010}. The hyperparameter $\beta$ tunes the degree of homogeneity of each region in the image. A small value of $\beta$ induces a noisy image with a large number of regions, contrary to a large value of $\beta$ that leads to few and large homogeneous regions. Finally, it is interesting to note that despite not knowing $C(\beta)$, drawing labels $\boldsymbol{z} = \left(z_1,\ldots,z_N\right)^T$ from the distribution \eqref{Potts} can be easily achieved by using a Gibbs sampler \cite{Robert}.

It is interesting to mention that while the Potts model is an effective means to introduce spatial correlation between discrete variables, there are other more complex models that could be investigated. In particular, Marroquin \emph{et al.} \cite{Marroquin2003} have shown that in segmentation applications better results may be obtained by using a two-layer hidden field, where hidden labels are assumed to be independent and correlation is introduced at a deeper layer by a vectorial Markov field. Similarly, Woolrich \emph{et al.} \cite{Woolrich2005} have proposed to approximate the Potts field by modeling mixture weights with a Gauss-Markov random field. However, these alternative models are not well adapted for $3$D images because they require significantly more computation and memory resources than the Potts model. These overheads result from the fact that they introduce $(K+1)N$ and $KN$ hidden variables respectively, against only $N$ for the Potts model ($N$ being the number of image pixels and $K$ the number of discrete states of the model).

\subsubsection{Parameter vector $\boldsymbol{\theta}$}
Assuming a priori independence between the parameters
$\boldsymbol{\theta}_{1},\ldots,\boldsymbol{\theta}_{K}$, the
joint prior for the parameter vector
$\boldsymbol{\theta}$
is
\begin{equation}\label{prior_theta}
  f\left(\boldsymbol{\theta}\right) =  \prod_{k=1}^K f(\boldsymbol{\theta}_{k})
\end{equation}
where $f(\boldsymbol{\theta}_{k})$ is the prior associated with the parameter vector $\boldsymbol{\theta_{k}}$ which mainly depends on the application considered. Two examples of priors $f\left(\boldsymbol{\theta}\right)$ will be investigated in Section \ref{sec:SimulationResults}.

\subsubsection{Granularity coefficient $\beta$}
As explained previously, fixing the value of $\beta$ a priori can be difficult because different images usually have different spatial organizations. A small value of $\beta$ will lead to a noisy classification and degrade the estimation of $\boldsymbol{\theta}$ and $\boldsymbol{z}$. Setting $\beta$ to a too large value will also degrade the estimation of $\boldsymbol{\theta}$ and $\boldsymbol{z}$ by producing over-smoothed classification results. Following a Bayesian approach, this paper proposes to assign $\beta$ an appropriate prior distribution and to estimate this coefficient jointly with $\left(\boldsymbol{\theta}, \boldsymbol{z}\right)$. In this work, the prior for the granularity coefficient $\beta$ is a uniform distribution on $(0,B)$
\begin{equation} \label{prior beta}
f(\beta) = \mathcal{U}_{(0,B)}(\beta)
\end{equation}
where $B = 2$ represents the maximum possible value of $\beta$. Note that it is unnecessary to consider larger values of $B$ since, for the first order neighborhood structure, ``when $\beta=2$, the Potts-Markov model is almost surely concentrated on single-color images'' \cite[p. 30]{Marin2007}.

\subsection{Posterior Distribution of $(\boldsymbol{\theta},\boldsymbol{z},\beta)$} \label{ssec:Posterior Distribution}
Assuming the unknown parameter vectors $\boldsymbol{\theta}$, $\boldsymbol{z}$, $\beta$ are a priori
independent and using Bayes theorem, the posterior distribution of $(\boldsymbol{\theta},\boldsymbol{z},\beta)$ can be expressed as
follows
\begin{equation} \label{posterior}
f\left(\boldsymbol{\theta},\boldsymbol{z},\beta|\boldsymbol{r}
\right) \propto
f(\boldsymbol{r}|\boldsymbol{\theta},\boldsymbol{z})f(\boldsymbol{\theta})f(\boldsymbol{z}|\beta)f(\beta)
\end{equation}
where $\propto$ means ``proportional to'' and where the likelihood
$f(\boldsymbol{r}|\boldsymbol{\theta},\boldsymbol{z})$ has been defined in \eqref{likelihood} and the
prior distributions $f(\boldsymbol{\theta})$, $f(\boldsymbol{z})$
and $f(\beta)$ in \eqref{prior_theta}, \eqref{Potts} and \eqref{prior beta}
respectively. Unfortunately the posterior distribution \eqref{posterior} is generally too complex to derive the MMSE or MAP estimators of the unknown parameters $\boldsymbol{\theta}$, $\boldsymbol{z}$ and $\beta$. One can think of using the EM algorithm to estimate these parameters. Indeed the EM algorithm has received much attention for mixture problems \cite{Dempster1977}. However, the shortcomings of the EM algorithm include \textit{convergence to local maxima or saddle points of the log-likelihood function and sensitivity to starting values} \cite[p. 259]{Diebolt1996}. An interesting alternative consists in using an MCMC method that generates samples that are asymptotically distributed according to the target distribution \eqref{posterior} \cite{Robert}. The generated samples are then used to approximate the Bayesian estimators. This strategy has been used successfully in several recent image processing applications (see \cite{Kayabol2009,Mignotte2010,Kayabol2010,Zhou2012,Destrempes2006,Nikou2010,Orieux2012} for examples in image filtering, dictionary learning, image reconstruction, fusion and segmentation). Many of these recent MCMC methods have been proposed for Bayesian models that include a Potts MRF \cite{Destrempes2006,Mignotte2007,Eches2010tgrs,Mignotte2010,PereyraTMIC2011}. However, these methods only studied the estimation of $\boldsymbol{\theta}$ and $\boldsymbol{z}$ conditionally to a known granularity coefficient $\beta$. The main contribution of this paper is to study Bayesian algorithms for the joint estimation of $\boldsymbol{\theta},\boldsymbol{z}$ and $\beta$. The next section studies a hybrid Gibbs sampler that generates samples that are asymptotically distributed according to the posterior \eqref{posterior}. The samples are then used to estimate the granularity coefficient $\beta$, the image labels $\boldsymbol{z}$ and the model parameter vector $\vartheta$. The resulting sampler can be easily adapted to existing MCMC algorithm where $\beta$ was previously assumed known, and can be applied to large images, both in $2$D and in $3$D.

\section{Hybrid Gibbs Sampler}\label{sec:Hybrid Gibbs Sampler}
This section studies a hybrid Metropolis-within-Gibbs sampler that generates samples that are asymptotically distributed according to
\eqref{posterior}. The conventional Gibbs sampler successively draws samples according to the full conditional distributions associated with the distribution of interest (here the posterior \eqref{posterior}). When a conditional distribution cannot be easily sampled, one can resort to a Metropolis-Hastings (MH) move, which generates samples according to an appropriate proposal and accept or reject these generated samples with a given probability. The resulting sampler is referred to as a Metropolis-within-Gibbs sampler (see \cite{Robert} for more details about MCMC methods). The sampler investigated in this section is based on the conditional distributions $\mathrm{P}[\boldsymbol{z}|\boldsymbol{\theta},\beta,\boldsymbol{r}]$, $f(\boldsymbol{\theta}|\boldsymbol{z},\beta,\boldsymbol{r})$ and $f(\beta|\boldsymbol{\theta},\boldsymbol{z},\boldsymbol{r})$ that are provided in the next paragraphs (see also Algorithm \ref{algo:hybridGibbs} below).

\begin{algorithm}
\caption{Proposed Hybrid Gibbs Sampler}
\label{algo:hybridGibbs}
    \begin{algorithmic}[1]
    \STATE Input: initial $\{\boldsymbol{\theta}^{(0)}, \boldsymbol{z}^{(0)}, \beta^{(0)}\}$, number of iterations $T$.
    \FOR{$t = 1$ to $T$}
    \STATE Generate $\boldsymbol{z}^{(t)} \sim \mathrm{P}[\boldsymbol{z}|\boldsymbol{\theta}^{(t-1)},\boldsymbol{z}^{(t-1)},\beta^{(t-1)},\boldsymbol{r}]$ according to \eqref{posteriorZ}
    \STATE Generate $\boldsymbol{\theta}^{(t)} \sim f(\boldsymbol{\theta}|\boldsymbol{\theta}^{(t-1)},\boldsymbol{z}^{(t)},\beta^{(t-1)},\boldsymbol{r})$ according to \eqref{posteriorTheta}
    \STATE Generate $\beta^{(t)} \sim f(\beta|\boldsymbol{\theta}^{(t)},\boldsymbol{z}^{(t)},\beta^{(t-1)},\boldsymbol{r})$ using Algorithm \ref{algo:ABC_LF_MH}.
    \ENDFOR
    \end{algorithmic}
\end{algorithm}

\subsection{Conditional probability $\mathrm{P}[\boldsymbol{z}|\boldsymbol{\theta},\beta,\boldsymbol{r}]$}\label{ssec:4.1}
For each voxel $n \in \{1,2,\ldots,N\}$, the class label $z_n$ is a
discrete random variable whose conditional distribution is fully
characterized by the probabilities
\begin{equation}
\mathrm{P}\left[z_n = k |
\boldsymbol{z}_{-n},\boldsymbol{\theta}_k,r_n,\beta\right] \propto
f(r_n|\boldsymbol{\theta}_k,z_n = k
)\mathrm{P}\left[z_n|\boldsymbol{z}_{\mathcal{V}(n)},\beta\right]
\end{equation}
where $k = 1,\ldots,K$, and where it is recalled that $\mathcal{V}(n)$ is the index set of the neighbors of the
$n$th voxel and $K$ is the number of classes. These probabilities can be expressed as
\begin{equation}\label{posteriorZUnnormalized}
\mathrm{P}\left[z_n = k | \boldsymbol{z}_{\mathcal{V}(n)}, \boldsymbol{\theta}_k, \beta, r_n \right] \propto \pi_{n,k}
\end{equation}
with
\begin{equation*}
\pi_{n,k} \triangleq \exp\left[\sum_{n'\in \mathcal{V}(n)} \beta \delta(k - z_{n'})\right] f(r_n|\boldsymbol{\theta}_k,z_n = k).
\end{equation*}
Once all the quantities $\pi_{n,k}$, $k=1,\ldots,K$, have been computed, they are normalized to obtain the probabilities $\tilde\pi_{n,k} \triangleq \mathrm{P}\left[z_n = k | \boldsymbol{z}_{\mathcal{V}(n)}, \boldsymbol{\theta}_k, \beta, r_n \right]$ as follows
\begin{equation}\label{posteriorZ}
 \tilde\pi_{n,k}= \frac{\pi_{n,k}}{\sum_{k=1}^K \pi_{n,k}}.
\end{equation}
Note that the probabilities of the label vector $\boldsymbol{z}$ in \eqref{posteriorZ} define an MRF. Sampling from this conditional distribution can be achieved by using a Gibbs sampler \cite{Robert} that draws discrete values in the finite set $\{1,\ldots,K\}$ with probabilities \eqref{posteriorZ}. More precisely, in this work $\boldsymbol{z}$ has been sampled using a $2$-color parallel chromatic Gibbs sampler that loops over $n \in \{1,2,\ldots,N\}$ following the checkerboard sequence \cite{Gonzalez2011}.

\subsection{Conditional probability density function $f(\boldsymbol{\theta}|\boldsymbol{z},\beta,\boldsymbol{r})$} \label{ssec:4.2}
The conditional density $f(\boldsymbol{\theta}|\boldsymbol{z},\beta,\boldsymbol{r})$ can be expressed as follows
\begin{equation}\label{posteriorTheta}
f(\boldsymbol{\theta}|\boldsymbol{z},\beta,\boldsymbol{r}) = f(\boldsymbol{\theta}|\boldsymbol{z},\boldsymbol{r}) \propto f(\boldsymbol{r}|\boldsymbol{\theta},\boldsymbol{z}) f(\boldsymbol{\theta})
\end{equation}
where $f(\boldsymbol{r}|\boldsymbol{\theta},\boldsymbol{z})$ and $f(\boldsymbol{\theta})$ have been defined in \eqref{likelihood} and \eqref{prior_theta}. Generating samples distributed according to \eqref{posteriorTheta} is strongly problem dependent. Some possibilities will be discussed in Sections \ref{sec:SimulationResults} and \ref{sec:Application to real data}. Generally, $\boldsymbol{\theta} =(\boldsymbol{\theta}_{1}^T,\ldots,\boldsymbol{\theta}_{K}^T)^T$ can be sampled coordinate-by-coordinate using the following Gibbs moves
\begin{equation}\label{posteriorTheta_k}
\boldsymbol{\theta}_k \sim f(\boldsymbol{\theta}_k|\boldsymbol{r},\boldsymbol{z}) \propto \prod_{\{n | z_n = k\}} f(r_n|\boldsymbol{\theta}_k)f(\boldsymbol{\theta}_k), \quad k=1,\ldots,K.
\end{equation}
In cases where sampling the conditional distribution \eqref{posteriorTheta_k} is too difficult, an MH move can be used resulting in a Metropolis-within-Gibbs sampler \cite{Robert} (details about the generation of samples $\boldsymbol{\theta}_k$ for the problems studied in Sections \ref{sec:SimulationResults} and \ref{sec:Application to real data} are provided in a separate technical report \cite{Pereyra_TIP_TechReport_2012}).

\subsection{Conditional probability density function $f(\beta|\boldsymbol{\theta},\boldsymbol{z},\boldsymbol{r})$} \label{ssec:beta}
From Bayes rule, the conditional density
$f(\beta|\boldsymbol{\theta},\boldsymbol{z},\boldsymbol{r})$ can be
expressed as follows
\begin{equation}\label{posteriorBeta}
f(\beta|\boldsymbol{\theta},\boldsymbol{z},\boldsymbol{r}) = f(\beta|\boldsymbol{z}) \propto f(\boldsymbol{z}|\beta)f(\beta)
\end{equation}
where $f(\boldsymbol{z}|\beta)$ and $f(\beta)$ have been defined in \eqref{Potts} and \eqref{prior beta} respectively. The generation of
samples according to $f(\beta|\boldsymbol{\theta},\boldsymbol{z},\boldsymbol{r})$ is not straightforward because $f(\boldsymbol{z}|\beta)$ is defined up to the unknown multiplicative constant $\frac{1}{C(\beta)}$ that depends on $\beta$. One could think of sampling $\beta$ by using an MH move, which requires computing the acceptance ratio
\begin{equation}\label{MH_beta}
\textrm{ratio} = \min\left\{1, \xi\right\}
\end{equation}
with \begin{equation}\label{ratio}
\xi =
\frac{f(\boldsymbol{z}|\beta^*)}{f(\boldsymbol{z}|\beta^{(t-1)})}
                \frac{f(\beta^*)}{f(\beta^{(t-1)})}
                \frac{q(\beta^{(t-1)}|\beta^*)}{q(\beta^*|\beta^{(t-1)})}
\end{equation}
where $\beta^* \sim q(\beta^*|\beta^{(t-1)})$ denotes an appropriate proposal distribution.
By replacing \eqref{Potts} into \eqref{ratio}, $\xi$ can be expressed as
\begin{equation} \label{ratio2}
\xi =  \frac{C(\beta^{(t-1)})}{C(\beta^*)} \frac{
\exp\left[\Phi_{\beta^*}(\boldsymbol{z})\right]}
                      {\exp\left[\Phi_{\beta^{(t-1)}}(\boldsymbol{z})\right]}
                \frac{f(\beta^*)}{f(\beta^{(t-1)})}
                \frac{q(\beta^{(t-1)}|\beta^*)}{q(\beta^*|\beta^{(t-1)})}
\end{equation}
where $\beta^*$ denotes the proposed value of $\beta$ at iteration $t$ and $\beta^{(t-1)}$ is the previous state of the chain. Unfortunately the ratio \eqref{ratio2} is generally intractable because of the term $\frac{C(\beta^{(t-1)})}{C(\beta^*)}$. The next section presents a likelihood-free MH algorithm that samples $\beta$ without requiring to evaluate $f(\boldsymbol{z}|\beta)$ and $C(\beta)$.

\section{Sampling the granularity coefficient}\label{subsec:auxiliary}
\subsection{Likelihood-free Metropolis-Hastings}
It has been shown in \cite{Marjoram2003} that it is possible to define a valid MH algorithm for posterior distributions with intractable likelihoods by introducing a carefully selected auxiliary variable and a tractable sufficient statistic on the target density. More precisely, consider an auxiliary vector $\boldsymbol{w}$ defined in the discrete state space $\left\{1,\ldots,K\right\}^{N}$ of $\boldsymbol{z}$ generated according to the likelihood $f(\boldsymbol{z}|\beta)$, i.e.,
\begin{equation} \label{ImpSampler2}
\boldsymbol{w} \sim f(\boldsymbol{w}|\beta) \triangleq
\frac{1}{C(\beta)}\exp\left[\Phi_{\beta}(\boldsymbol{w})\right]
\end{equation}
Also, let $\eta(\boldsymbol{z})$ be a tractable sufficient statistic of $\boldsymbol{z}$, i.e., $f(\beta | \boldsymbol{z}) = f[\beta | \eta(\boldsymbol{z})]$. Then, it is possible to generate samples that are asymptotically distributed according to the exact conditional density $f(\beta|\boldsymbol{\theta},\boldsymbol{z},\boldsymbol{r})=f(\beta|\boldsymbol{z})$ by introducing an additional rejection step based on $\eta(\boldsymbol{z})$ into a standard MH move \cite{Marjoram2003} (see Algorithm \ref{algo:Exact_LF_MH} below).
\begin{algorithm}
\caption{Exact likelihood-free MH step \cite{Marjoram2003}}
\label{algo:Exact_LF_MH}
    \begin{algorithmic}[1]
    \STATE Input: $\{\beta^{(t-1)},\boldsymbol{z}^{(t)}\}$
    \STATE Generate $\beta^* \sim q\left(\beta^* |\beta^{(t-1)}\right)$
    \STATE Generate an auxiliary variable $\boldsymbol{w} \sim f(\boldsymbol{w}|\beta^*)$
    \IF {$\eta(\boldsymbol{w}) = \eta(\boldsymbol{z}^{(t)})$}
        \STATE Set $\textrm{ratio} = \frac{f(\beta^*)}{f(\beta^{(t-1)})}
                \frac{q(\beta^{(t-1)}|\beta^*)}{q(\beta^*|\beta^{(t-1)})}$
        \STATE Draw $u \sim \mathcal{U}_{(0, 1)}$
        \IF {$(u < \textrm{ratio})$}
            \STATE Set $\beta^{(t)} = \beta^*$
        \ELSE
            \STATE Set $\beta^{(t)} = \beta^{(t-1)}$
        \ENDIF
    \ELSE
        \STATE Set $\beta^{(t)} = \beta^{(t-1)}$
    \ENDIF
    \end{algorithmic}
\end{algorithm}

Note that the MH acceptance ratio in algorithm \ref{algo:Exact_LF_MH} is the product of the prior ratio $\frac{f(\beta^*)}{f(\beta^{(t-1)})}$ and the proposal ratio $\frac{q(\beta^{(t-1)}|\beta^*)}{q(\beta^*|\beta^{(t-1)})}$. The generally intractable likelihood ratio $\frac{f(\boldsymbol{z}|\beta^*)}{f(\boldsymbol{z}|\beta^{(t-1)})}$ has been replaced by the simulation and rejection steps involving the discrete auxiliary vector $\boldsymbol{w}$. Despite not computing $\frac{f(\boldsymbol{z}|\beta^*)}{f(\boldsymbol{z}|\beta^{(t-1)})}$ explicitly, the resulting MH move still accepts candidate values $\beta^*$ with the correct probability \eqref{MH_beta} \cite{Marjoram2003}.

Unfortunately exact likelihood-free MH algorithms have several shortcomings \cite{Marin2011}. For instance,
their acceptance ratio is generally very low because candidates $\beta^*$ are only accepted if they lead to an auxiliary
vector $\boldsymbol{w}$ that verifies $\eta(\boldsymbol{z}^{(t)}) = \eta(\boldsymbol{w})$. In addition, most Bayesian models do not
have known sufficient statistics. These limitations have been addressed in the ABC framework by introducing an approximate
likelihood-free MH algorithm (henceforth denoted as ABC-MH) \cite{Marjoram2003}. Precisely, the ABC-MH algorithm does not require the
use of a sufficient statistic and is defined by a less restrictive criterion of the form $\rho\left[\eta(\boldsymbol{z}^{(t)}), \eta(\boldsymbol{w})\right] < \epsilon$, where $\rho$ is an arbitrary distance measure and
$\epsilon$ is a tolerance parameter (note that this criterion can be applied to both discrete and continuous intractable distributions, contrary to algorithm \ref{algo:Exact_LF_MH} that can only be applied to discrete distribution). The resulting algorithm generates samples that are asymptotically distributed according to an
approximate posterior density \cite{Marjoram2003}
\begin{equation}\label{approx_posterior_beta}
f_\epsilon(\beta|\boldsymbol{z}) \approx \sum_{\boldsymbol{w}} f(\beta) f(\boldsymbol{w}| \beta) \boldsymbol{1}_{\left[\rho\left[\eta(\boldsymbol{z}), \eta(\boldsymbol{w})\right] < \epsilon\right]}(\boldsymbol{w})
\end{equation}
whose accuracy depends on the choice of $\eta(\boldsymbol{z})$ and $\epsilon$ (if $\eta(\boldsymbol{z})$ is a sufficient statistic and $\epsilon = 0$, then \eqref{approx_posterior_beta} corresponds to the exact posterior density).

In addition, note that in the exact likelihood-free MH algorithm, the auxiliary vector $\boldsymbol{w}$ has to be generated using perfect sampling \cite{Propp1996,Childs2001}. This constitutes a major limitation, since perfect or exact sampling techniques \cite{Propp1996,Childs2001} are too costly for image processing applications where the dimension of $\boldsymbol{z}$ and $\boldsymbol{w}$ can exceed one million pixels. A convenient alternative is to replace perfect simulation by a few Gibbs moves with target density $f(\boldsymbol{w}|\beta^*)$ as proposed in \cite{Grelaud2009}. The accuracy of this second approximation depends on the number of moves and on the initial state of the sampler. An infinite number of moves would clearly lead to perfect simulation regardless of the initialization. Inspired from \cite{Liang2010}, we propose to use $\boldsymbol{z}$ as initial state to produce a good approximation with a small number of moves. A simple explanation for this choice is that for candidates $\beta^*$ close to the mode of $f(\beta|\boldsymbol{z})$, the vector $\boldsymbol{z}$ has a high likelihood $f(\boldsymbol{z}|\beta)$. In other terms, using $\boldsymbol{z}$ as initial state does not lead to perfect sampling but provides a good final approximation of $f(\beta|\boldsymbol{z})$ around its mode. The accuracy of this approximation can be easily improved by increasing the number of moves at the expense of computing time. However, several simulation results in \cite{Pereyra_TIP_TechReport_2012,Everitt2012} have shown that the resulting ABC algorithm approximates $f(\beta | \boldsymbol{z})$ correctly even for a small number of moves.

\subsection{Choice of $\eta(\boldsymbol{z})$, $\rho$ and $\epsilon$}
As explained previously, ABC algorithms require defining an appropriate statistic $\eta(\boldsymbol{z})$, a distance function $\rho$ and a tolerance level $\epsilon$. The choice of $\eta(\boldsymbol{z})$ and $\rho$ are fundamental to the success of the approximation, while the value of $\epsilon$ is generally less important \cite{Marin2011}. Fortunately the Potts MRF, being a Gibbs random field, belongs to the exponential family and has the following one-dimensional sufficient statistic \cite{Grelaud2009,Marin2011}
\begin{equation}\label{statistic}
\eta(\boldsymbol{z}) \triangleq \sum_{n=1}^N\sum_{n'\in \mathcal{V}(n)} \delta (z_{n} - z_{n'})
\end{equation}
where it is recalled that $\mathcal{V}(n)$ is the index set of the neighbors of the $n$th voxel. Note that because \eqref{statistic} is a sufficient statistic, the approximate posterior $f_\epsilon(\beta|\boldsymbol{z})$ tends to the exact posterior $f(\beta|\boldsymbol{z})$ as $\epsilon \rightarrow 0$ \cite{Marjoram2003}.

The distance function $\rho$ considered in this work is the one-dimensional Euclidean distance
\begin{equation}\label{rho}
\rho\left[\eta(\boldsymbol{z}), \eta(\boldsymbol{w})\right] = |\eta(\boldsymbol{z}) -  \eta(\boldsymbol{w})|
\end{equation}
which is a standard choice in ABC methods \cite{Marin2011}. Note from \eqref{statistic} and \eqref{rho} that the distance $\rho[\cdot,\cdot]$ between $\eta(\boldsymbol{z})$ and $\eta(\boldsymbol{w})$ reduces to the difference in the number of active cliques in $\boldsymbol{z}$ and $\boldsymbol{w}$. It is then natural to set the tolerance as a fraction of that number, i.e., $\epsilon = \nu \eta(\boldsymbol{z})$ ($\nu = \frac{1}{1000}$ will be used in our experiments). Note that the choice of $\nu$ is crucial when the prior density $f(\beta)$ is informative because increasing $\nu$ introduces estimation bias by allowing the posterior density to drift towards the prior \cite{Beaumont2002}. However, in this work the choice of $\nu$ is less critical because $\beta$ has been assigned a flat prior distribution.

\subsection{Proposal distribution $q(\beta^* | \beta^{(t-1)})$}
Finally, the proposal distribution $q(\beta^* | \beta^{(t-1)})$ used to explore the set $(0,B)$ is chosen as a truncated normal distribution centered on the previous value of the chain with variance $s^2_\beta$
\begin{equation} \label{proposalBeta}
\beta^* \sim \mathcal{N}_{(0,B)}\left(\beta^{(t-1)},s^2_\beta \right).
\end{equation}
where the variance $s^2_\beta$ is adjusted during the \emph{burn-in} period to ensure an acceptance ratio close to $5\%$, as recommended in \cite{Pereyra_TIP_TechReport_2012}. This proposal strategy is referred to as random walk MH algorithm \cite[p. 245]{Robert}. The choice of this proposal distribution has been motivated by the fact that for medium and large problems (i.e., Markov fields larger than $50\times50$ pixels) the distribution $f(\beta |\boldsymbol{z})$ becomes very sharp and can be efficiently explored using a random walk.

The resulting ABC MH method is summarized in Algorithm \ref{algo:ABC_LF_MH} below.
Note that Algorithm \ref{algo:ABC_LF_MH} corresponds to step $5$ in Algorithm \ref{algo:hybridGibbs}.
\begin{algorithm}
\caption{ABC likelihood-free MH step \cite{Marjoram2003}}
\label{algo:ABC_LF_MH}
    \begin{algorithmic}[1]
    \STATE Input: $\{\beta^{(t-1)},\boldsymbol{z}^{(t)},\nu, s^2_\beta \}$, number of moves $M$.
    \STATE Generate $\beta^* \sim \mathcal{N}_{(0,B)}\left(\beta^{(t-1)},s^2_\beta\right)$
    \STATE Generate $\boldsymbol{w} \sim f(\boldsymbol{w}|\beta^*)$ through $M$ Gibbs moves with initial state $\boldsymbol{z}^{(t)}$
    \IF {$|\eta(\boldsymbol{z}^{(t)})-\eta(\boldsymbol{w})|<\nu \eta(\boldsymbol{z}^{(t)})$}
        \STATE Set $\textrm{ratio} = \frac{f(\beta^*)}{f(\beta^{(t-1)})}
                \frac{q(\beta^{(t-1)}|\beta^*)}{q(\beta^*|\beta^{(t-1)})}$
        \STATE Draw $u \sim \mathcal{U}_{(0, 1)}$
        \IF {$(u < \textrm{ratio})$}
            \STATE Set $\beta^{(t)} = \beta^*$
        \ELSE
            \STATE Set $\beta^{(t)} = \beta^{(t-1)}$
        \ENDIF
    \ELSE
        \STATE Set $\beta^{(t)} = \beta^{(t-1)}$
    \ENDIF
    \end{algorithmic}
\end{algorithm}

\section{Experiments} \label{sec:SimulationResults}
This section presents simulation results conducted on synthetic data to assess the importance of estimating the hyperparameter $\beta$ from data as opposed to fixing it a priori (i.e., the advantage of estimating the posterior $p(\boldsymbol{\theta},\boldsymbol{z},\beta|\boldsymbol{r})$ instead of fixing $\beta$). Simulations have been performed as follows: label vectors distributed according to a Potts MRF have been generated using different granularity coefficients (in this work bidimensional fields of size $256 \times 256$ pixels have been considered). Each label vector has in turn been used to generate an observation vector following the observation model \eqref{mixture}. Finally, samples distributed according to the posterior distribution of the unknown parameters $(\boldsymbol{\theta},\boldsymbol{z},\beta)$ have been estimated from each observation vector using Algorithm \ref{algo:hybridGibbs} coupled with Algorithm \ref{algo:ABC_LF_MH} (assuming the number of classes $K$ is known). The performance of the proposed algorithm has been assessed by comparing the resulting Bayesian estimates with the true values of the parameters. This paper presents simulation results obtained using two different mixture models. Additional simulation results using other mixture models are available in a separate technical report \cite{Pereyra_TIP_TechReport_2012}. Finally, comparisons with the state-of-the-art methods proposed in \cite{Moller2006,Murray2006,Risser2010} are also reported in \cite{Pereyra_TIP_TechReport_2012}.


\subsection{Mixture of gamma distributions} \label{ssec:mixture_gamma}
The first experiment considers a mixture of gamma distributions. This observation model is frequently used to describe the statistics of pixels in multilook SAR images and has been extensively applied for SAR image segmentation \cite{Tourneret2003}. Accordingly, the conditional observation model \eqref{mixture} is defined by a gamma distribution with parameters $L$ and $m_k$ \cite{Tourneret2003}
\begin{equation}
r_n | z_n = k \sim f(r_n | \boldsymbol{\theta}_k) = \left(\frac{L}{m_k}\right)^L \frac{r_n^{L-1}}{\Gamma(L)}\exp\left(-\frac{L r_n}{m_k}\right)
\end{equation}\label{eq:pGammaL} where $\Gamma(t) = \int_0^{+\infty} u^{t-1} e^{-u} du$ is the standard Gamma function and $L$ (the number of looks) is assumed to be known ($L = 3$ in this paper). The means $m_k$ ($k = 1,\ldots,K$) are assigned inverse gamma prior distributions as in \cite{Tourneret2003}. The estimation of $\beta$, $\boldsymbol{z}$ and $\boldsymbol{\theta} = \boldsymbol{m}=(m_1,\ldots,m_K)^T$ is then achieved by using Algorithm \ref{algo:hybridGibbs}. The sampling strategies described in Sections \ref{ssec:4.1} and \ref{subsec:auxiliary} can be used for the generation of samples according to $P[\boldsymbol{z}| \boldsymbol{m},\beta,\boldsymbol{r}]$ and $f(\beta| \boldsymbol{m},\boldsymbol{z},\boldsymbol{r})$. More details about simulation according to $f(\boldsymbol{m}|\boldsymbol{z},\beta, \boldsymbol{r})$ are provided in the technical report \cite{Pereyra_TIP_TechReport_2012}.

\begin{figure}[htb]
\begin{minipage}[a1]{.49\linewidth}
  \centering
  \centerline{\includegraphics[width=4.5cm]{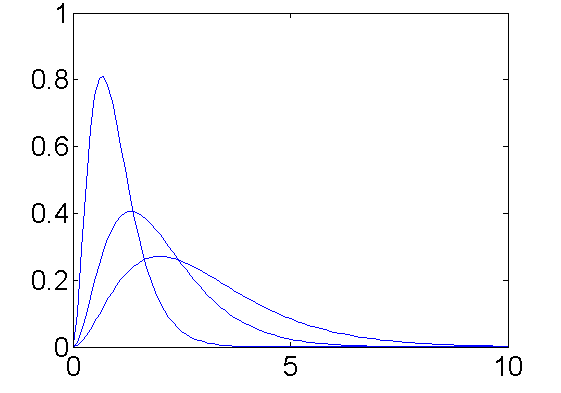}}
  \centerline{(a)  \small{gamma mixture}}\medskip
\end{minipage}
\begin{minipage}[a1]{.49\linewidth}
  \centering
  \centerline{\includegraphics[width=4.5cm]{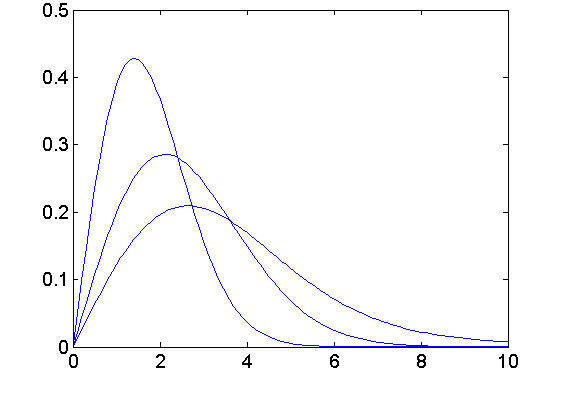}}
  \centerline{(b)  \small{$\alpha$-Rayleigh mixture}}\medskip
\end{minipage}
\caption{\small{Probability density functions of the distributions mixed for the first set and the second set of experiments.}}
\label{fig:Gamma_aRayleigh_mixture}
\end{figure}

\begin{figure}[htb]
\begin{minipage}[a1]{.49\linewidth}
  \centering
  \centerline{\includegraphics[width=4.0cm]{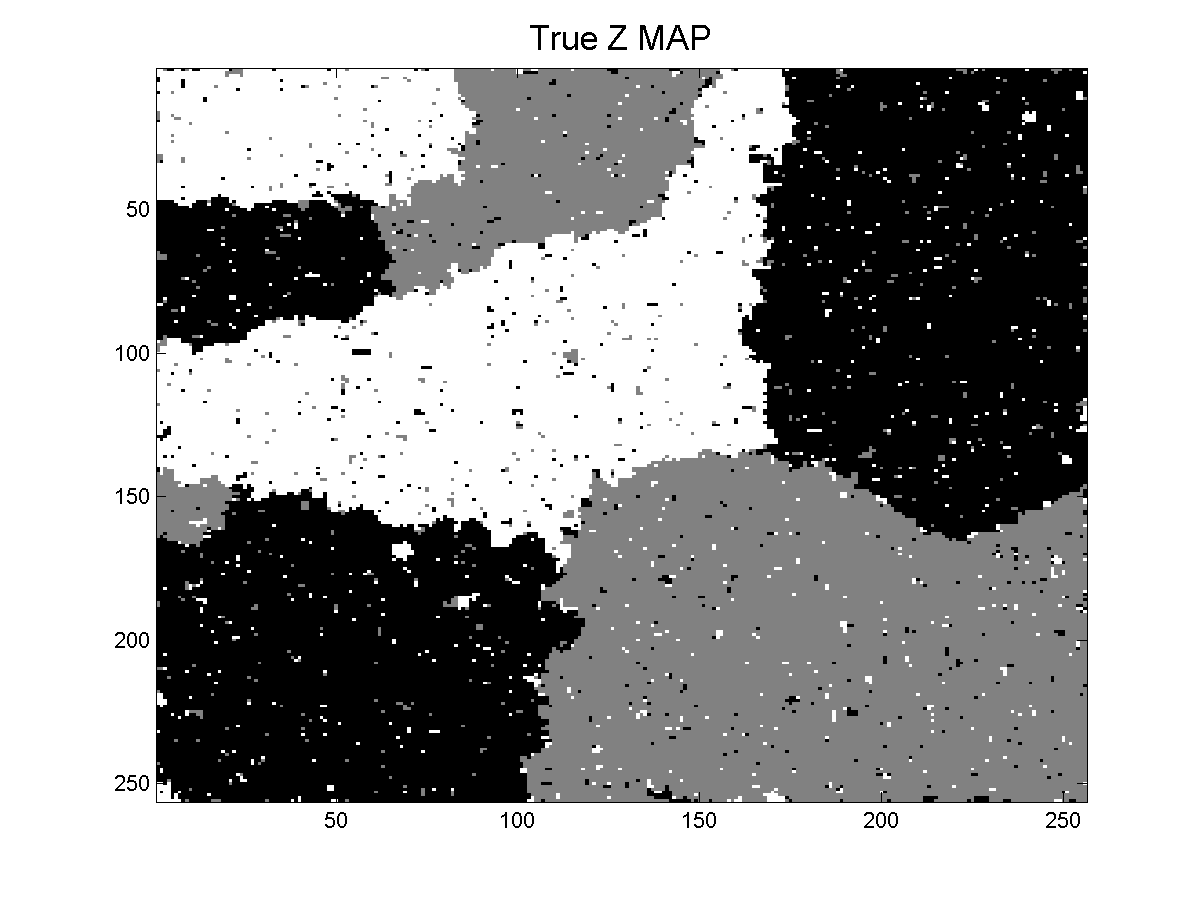}}
  \centerline{(a) \small{True Labels ($\beta = 1.2$)}}\medskip
\end{minipage}
\hfil
\begin{minipage}[a1]{.49\linewidth}
  \centering
  \centerline{\includegraphics[width=4.0cm]{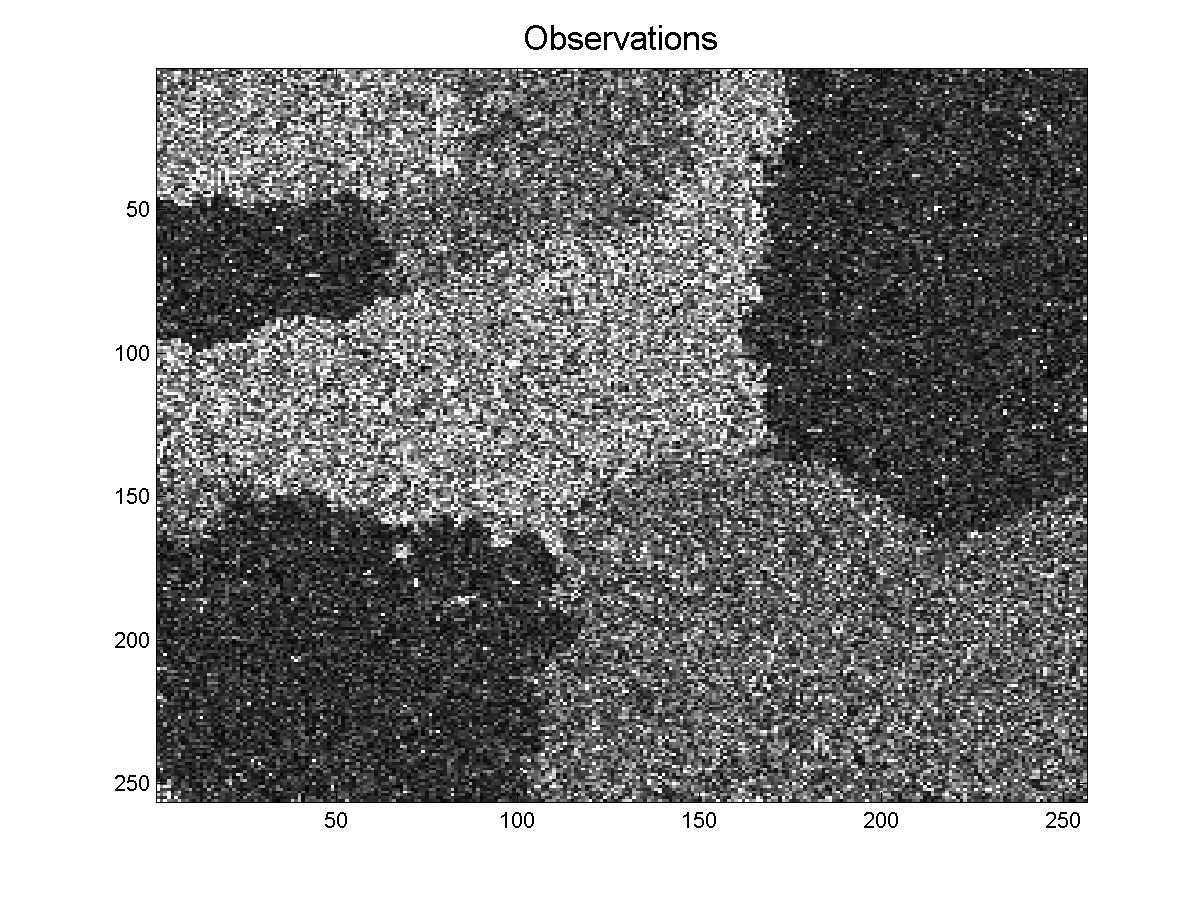}}
  \centerline{(b)  \small{Observations}}\medskip
\end{minipage}
\hfill
\begin{minipage}[a1]{.49\linewidth}
  \centering
  \centerline{\includegraphics[width=4.0cm]{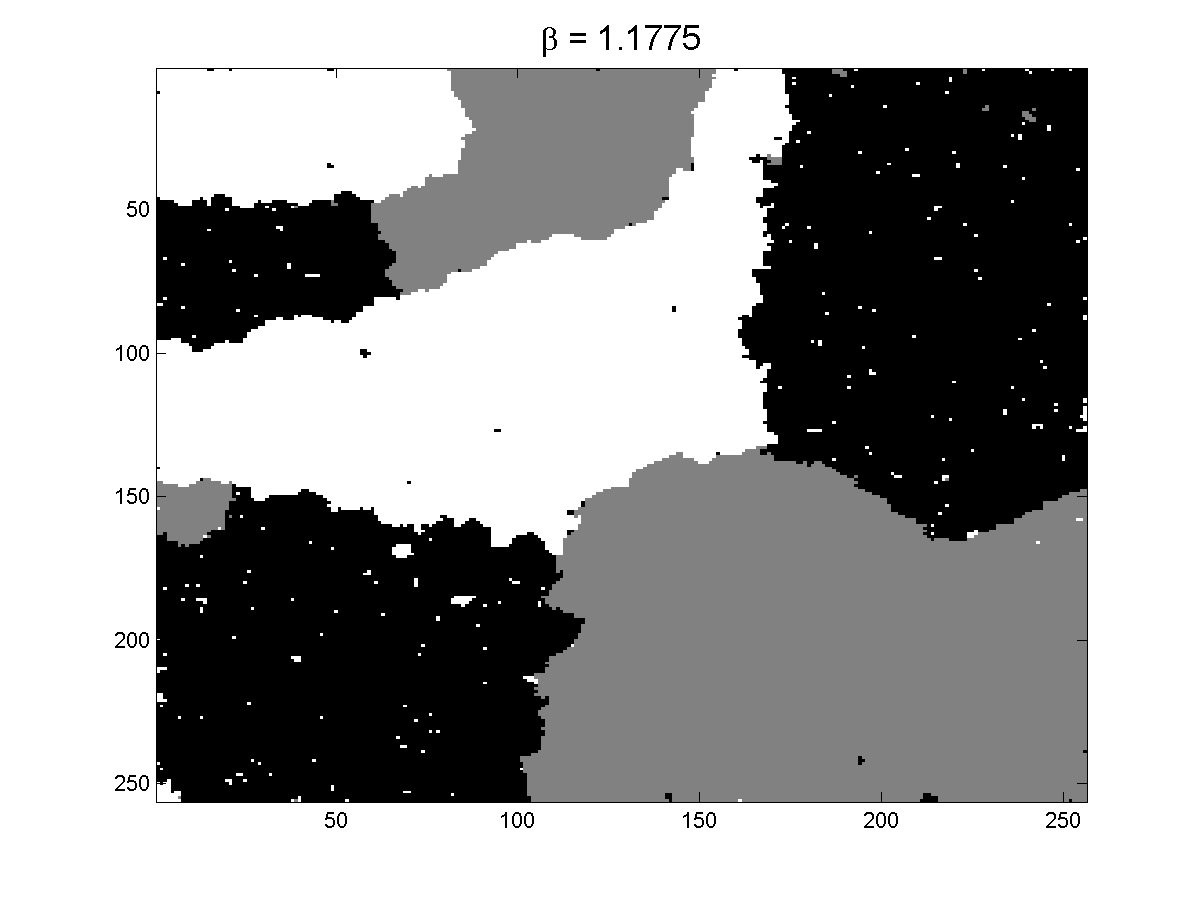}}
  \centerline{(c)  \small{Estimated $\beta$}}\medskip
\end{minipage}
\hfill
\begin{minipage}[a2]{.49\linewidth}
  \centering
  \centerline{\includegraphics[width=4.0cm]{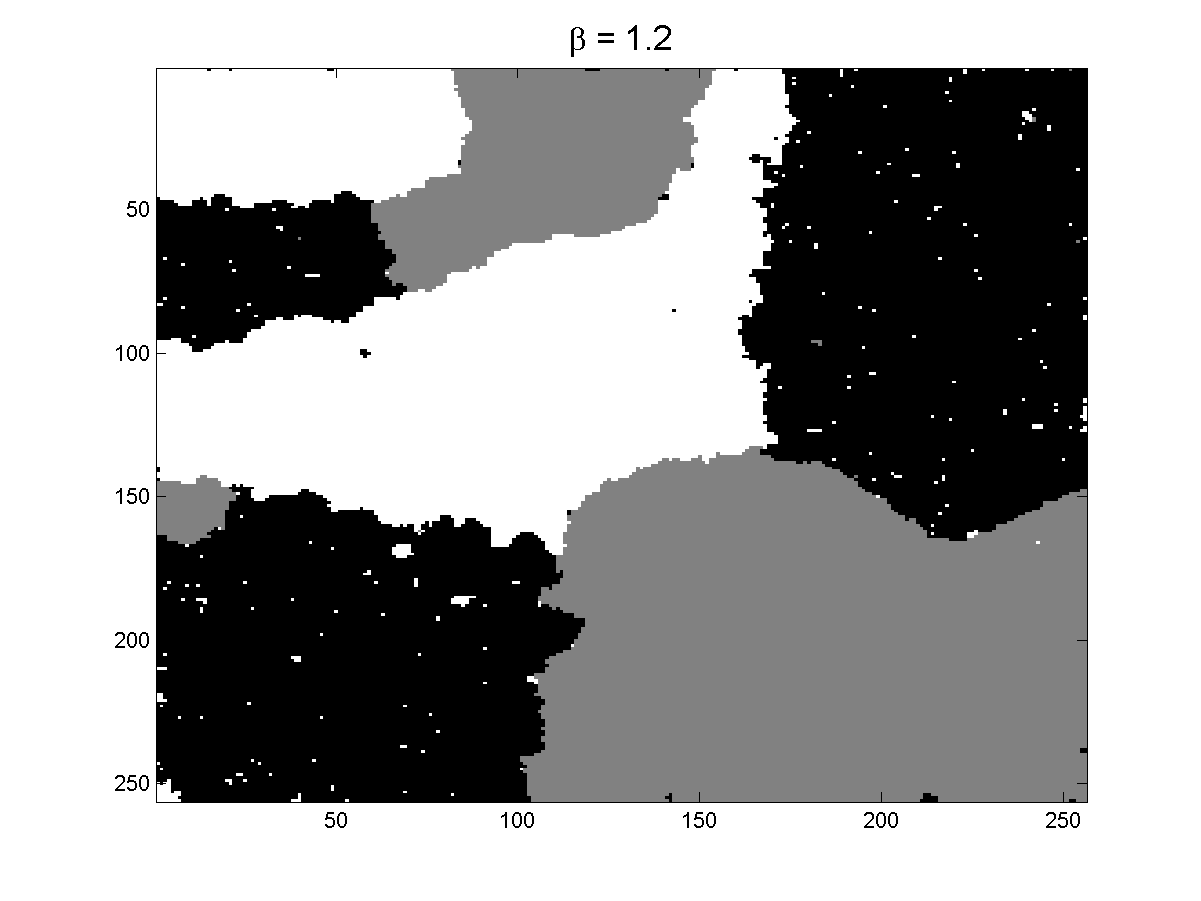}}
  \centerline{(d)  \small{True $\beta = 1.2$}}\medskip
\end{minipage}

\begin{minipage}[a2]{.49\linewidth}
  \centering
  \centerline{\includegraphics[width=4.0cm]{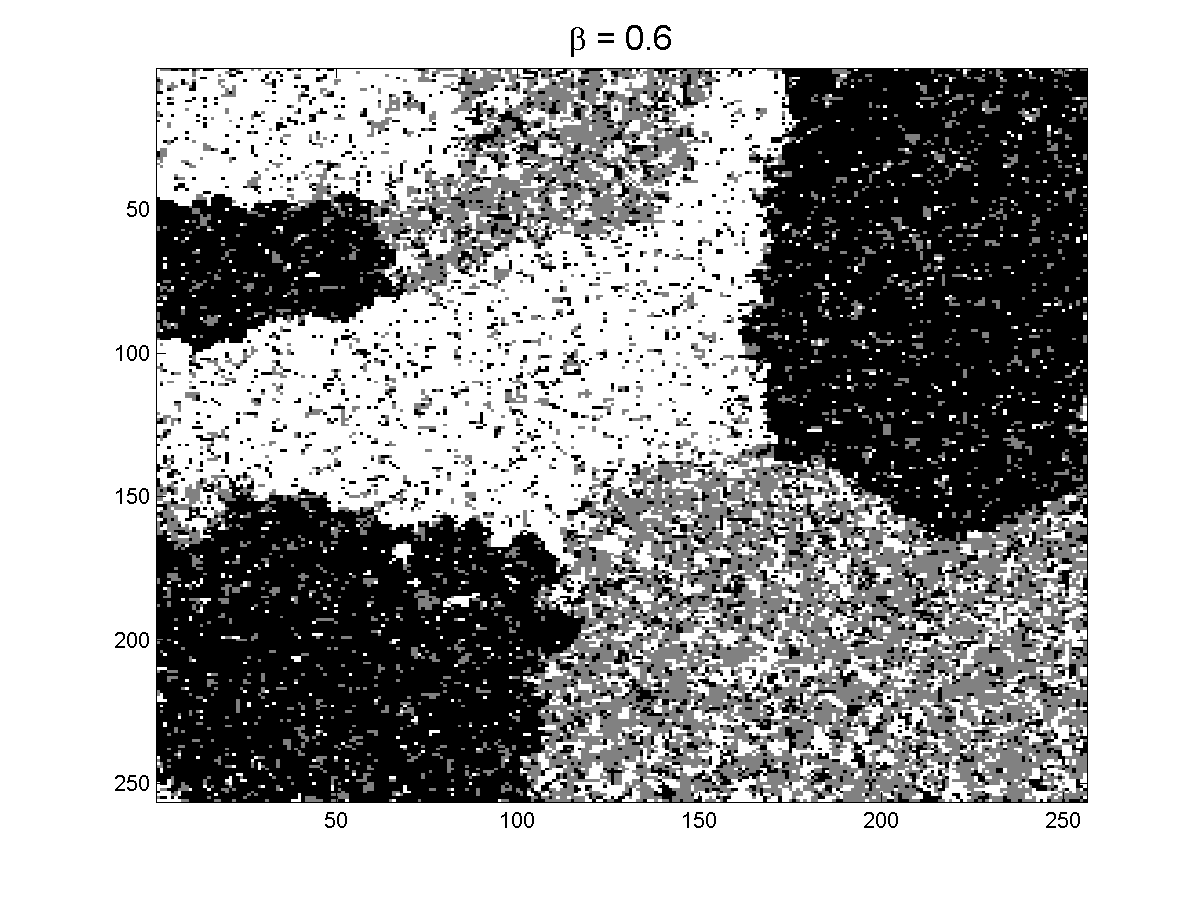}}
  \centerline{(e)  \small{$\beta = 0.6$}}\medskip
\end{minipage}
\hfill
\begin{minipage}[a2]{.49\linewidth}
  \centering
  \centerline{\includegraphics[width=4.0cm]{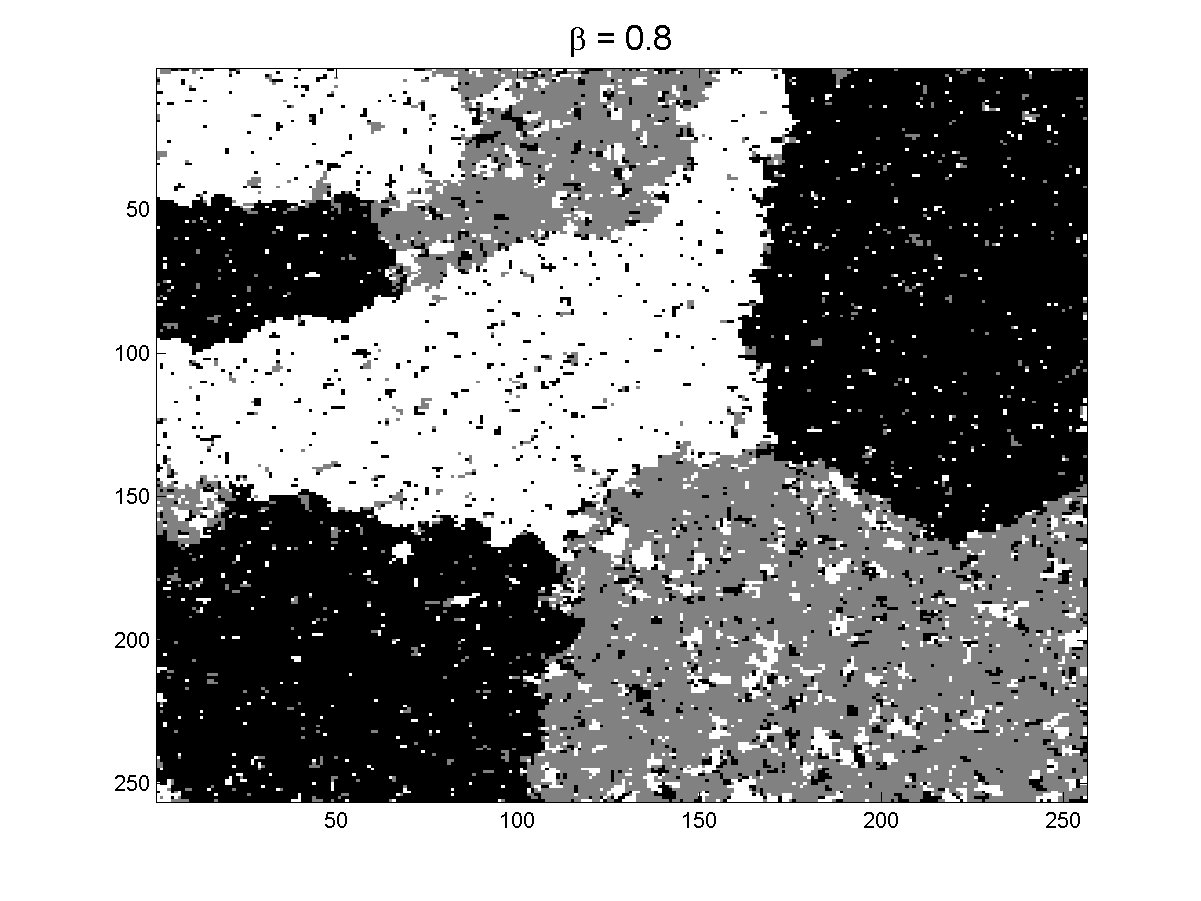}}
  \centerline{(f) \small{$\beta = 0.8$}}\medskip
\end{minipage}
\hfill
\begin{minipage}[a2]{.49\linewidth}
  \centering
  \centerline{\includegraphics[width=4.0cm]{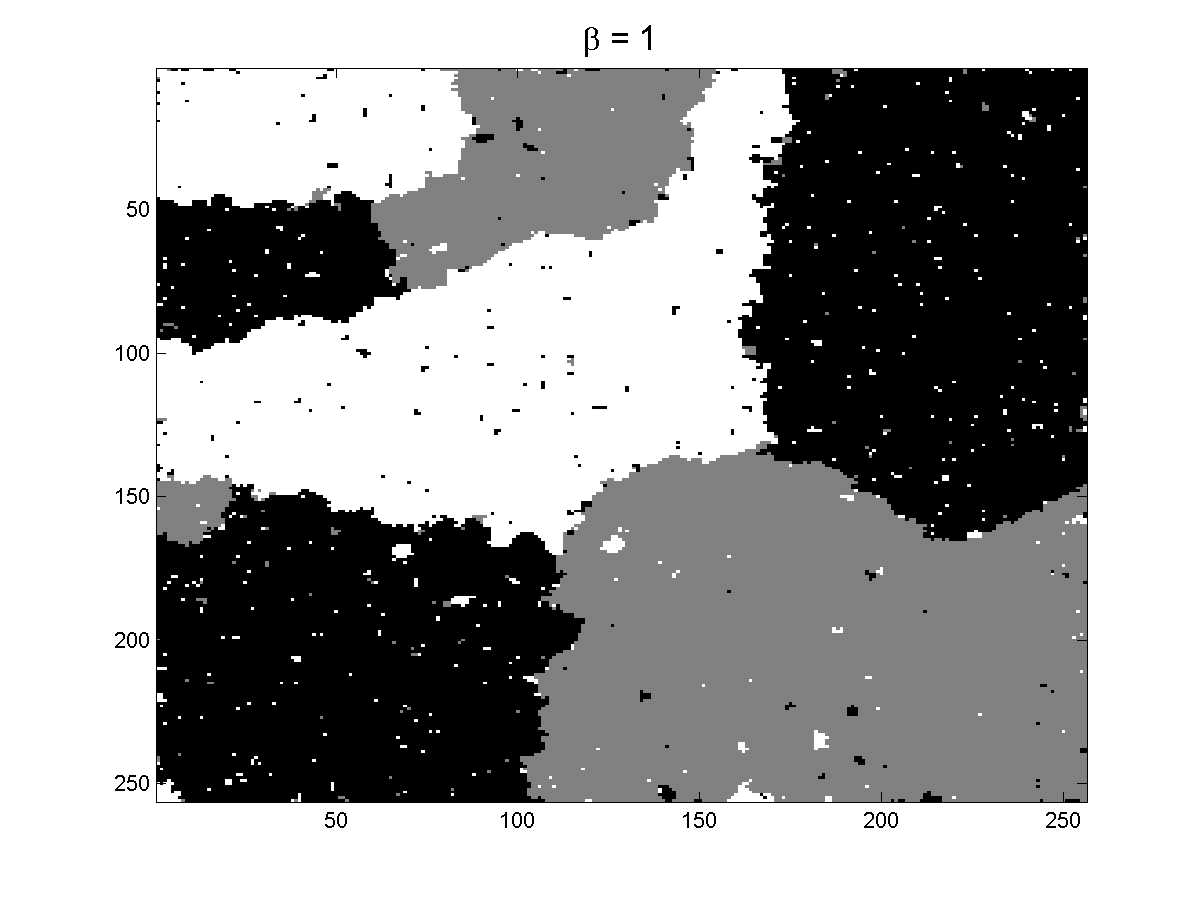}}
  \centerline{(g)  \small{$\beta = 1.0$}}\medskip
\end{minipage}
\hfill
\begin{minipage}[a2]{.49\linewidth}
  \centering
  \centerline{\includegraphics[width=4.0cm]{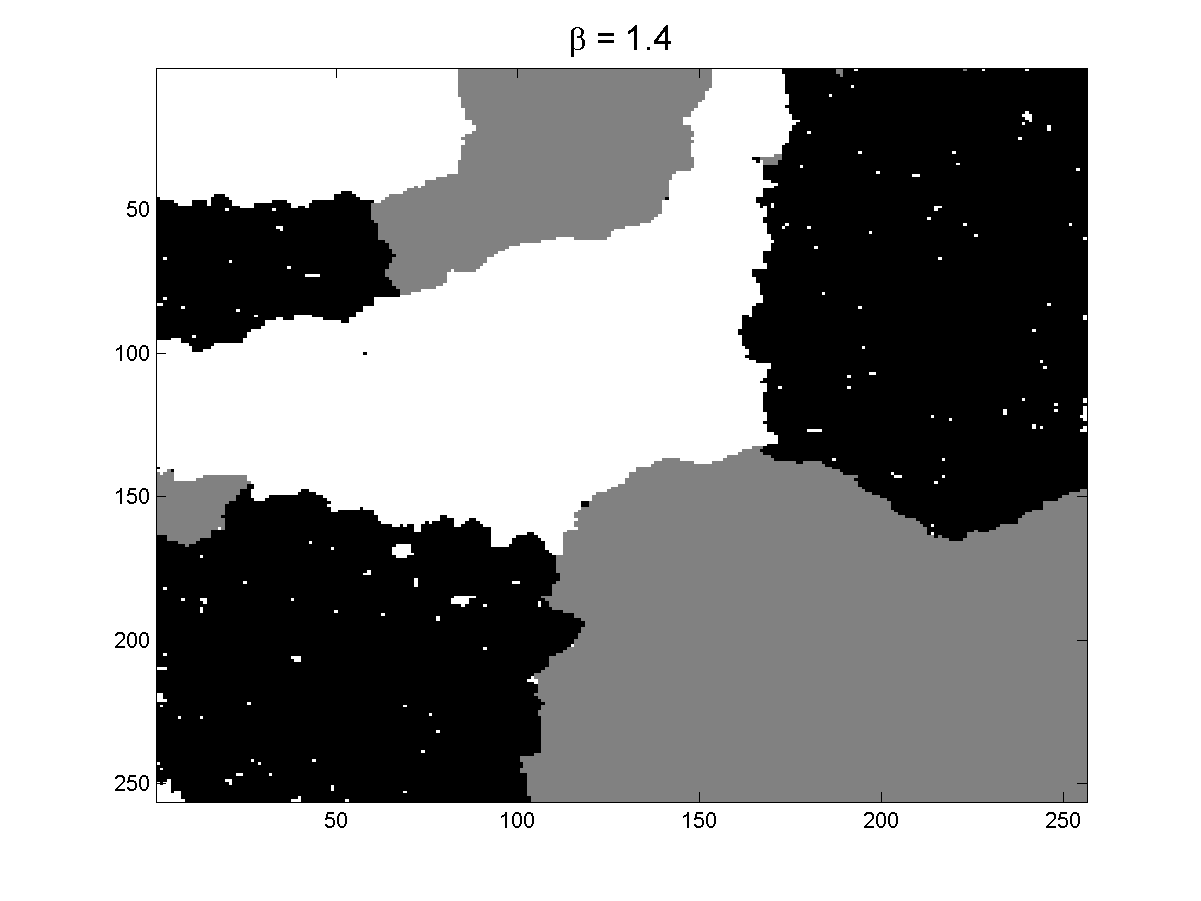}}
  \centerline{(h)  \small{$\beta = 1.4$}}\medskip
\end{minipage}
\caption{\small{Gamma mixture: estimated labels using the MAP estimators. (a) Ground truth, (b) observations, (c) proposed algorithm (estimated $\beta$),(d) true $\beta = 1.2$, (e)-(h) fixed $\beta$ = (0.6, 0.8, 1.0, 1.2, 1.4).}}
\label{fig:betaPlot3}
\end{figure}

The first results have been obtained for a $3$-component gamma mixture with parameters $\boldsymbol{m} = (1;2;3)$. Fig. \ref{fig:Gamma_aRayleigh_mixture}(a) shows the densities of the gamma distributions defining the mixture model. Note that there is significant overlap between the densities making the inference problem very challenging. For each experiment the MAP estimates of the class labels $\boldsymbol{z}$ have been computed from a single Markov chain of $T=1\,000$ iterations whose first $400$ iterations (burn-in period) have been removed. Table \ref{table6} shows the percentage of MAP class labels correctly estimated. The first column corresponds to labels that were estimated jointly with $\beta$ whereas the other columns result from fixing $\beta$ to different a priori values. To ease interpretation, the best and second best results for each simulation scenario in Table \ref{table6} are highlighted in red and blue. We observe that the proposed method performs as well as if $\beta$ was perfectly known. On the other hand, setting $\beta$ to an incorrect value may severely degrade estimation performance. Table \ref{table6b} shows the MMSE estimates of $\beta$ and $\boldsymbol{m}$ corresponding to the three simulations of the first column of Table \ref{table6} (proposed method) as well as the standard deviations of the estimates (results are displayed as [mean $\pm$ standard deviation]). We observe that these values are in good agreement with the true values used to generate the observation vectors. Finally, for illustration purposes, Fig. \ref{fig:betaPlot3} shows the MAP estimates of the class labels corresponding to the simulation scenario reported in the last row of Table \ref{table6}. More precisely, Fig. \ref{fig:betaPlot3}(a) depicts the class label map, which is a realization of a 3-class Potts MRF with $\beta = 1.2$. The corresponding synthetic image is presented in Fig. \ref{fig:betaPlot3}(b). Fig. \ref{fig:betaPlot3}(c) shows the class labels obtained with the proposed method and Fig. \ref{fig:betaPlot3}(d) those obtained when $\beta$ is perfectly known. Lastly, Figs. \ref{fig:betaPlot3}(e)-(h) show the results obtained when $\beta$ is fixed incorrectly to $0.6,0.8,1.0$ and $1.4$. We observe that the classification produced by the proposed method is very close to that obtained by fixing $\beta$ to its true value, whereas fixing $\beta$ incorrectly results in either noisy or excessively smooth results.

\begin{table}[htb]
\renewcommand{\arraystretch}{1.3}
\caption{Gamma Mixture: Class label estimation ($K=3$)}
\label{table6}
\centering
\tabcolsep 5.0pt
\tiny
\begin{tabular}{|c||c|c|c|c|c|c|c|}
  \cline{4-8}
  \multicolumn{3}{c|}{} & \multicolumn{5}{c|}{Correct classification with $\beta$ fixed} \\
  \cline{2-8}
  \multicolumn{1}{c|}{} & \multicolumn{2}{c|}{Proposed method} & $\beta=0.6$ & $\beta=0.8$ & $\beta=1.0$ & $\beta=1.2$ & $\beta=1.4$ \\
  \hline
  \hline
  $\textrm{True } \beta = 0.8$ & $\hat{\beta}=0.80$ & \red62.2\% & 61.6\% & \blue61.7\% & 58.8\% & 41.5\% & 40.1\% \\
  \hline
  $\textrm{True } \beta = 1.0$ & $\hat{\beta}=1.00$ &\red77.9\% & 67.3\% & 73.4\% & \blue77.7\% & 75.9\% & 74.2\% \\
  \hline
  $\textrm{True } \beta = 1.2$ & $\hat{\beta}=1.18$ &\red95.6\% & 76.6\% & 87.8\% & 94.9\% & \red95.6\% & 95.5\% \\
  \hline
\end{tabular}
\end{table}

\begin{table}[htb]
\renewcommand{\arraystretch}{1.3}
\caption{Gamma Mixture: Parameter estimation}
\label{table6b}
\centering
\tabcolsep 5.0pt
\tiny
\begin{tabular}{|c||c|c||c|c||c|c|}
  \cline{2-7}
  \multicolumn{1}{c|}{} & true & MMSE & true & MMSE & true & MMSE \\
  \hline
  \hline
  $\beta$      & 0.80  & 0.80 $\pm$ 0.01    & 1.00 & 1.00 $\pm$ 0.01 & 1.20 & 1.18 $\pm$ 0.02 \\
  \hline
  $m_1$       & 1     & 0.99 $\pm$ 0.02    & 1     & 1.00 $\pm$ 0.02 & 1   & 0.99 $\pm$ 0.03 \\
  \hline
  $m_2$       & 2     & 1.99 $\pm$ 0.02    & 2     & 1.98 $\pm$ 0.02 & 2   & 1.98 $\pm$ 0.07 \\
  \hline
  $m_3$       & 3     & 2.98 $\pm$ 0.03    & 3     & 2.98 $\pm$ 0.04 & 3   & 3.01 $\pm$ 0.03 \\
  \hline
\end{tabular}
\end{table}

\subsection{Mixture of $\alpha$-Rayleigh distributions}
The second set of experiments has been conducted using a mixture of $\alpha$-Rayleigh distributions. This observation model has been recently proposed to describe ultrasound images of dermis \cite{PereyraUFFC2011} and has been successfully applied to the segmentation of skin lesions in $3$D ultrasound images \cite{PereyraTMIC2011}. Accordingly, the conditional observation model \eqref{mixture} used in the experiments is defined by an $\alpha$-Rayleigh distribution
\begin{equation}
r_n|z_n = k \sim f(r_n|\boldsymbol{\theta}_k) = p_{\alpha\mathcal{R}}(r_n|\alpha_k,\gamma_k) \label{eq:palphaR}
\end{equation}
with
\begin{equation*}
p_{\alpha\mathcal{R}}(r_n|\alpha_k,\gamma_k) \triangleq r_{n}
\int_{0}^{\infty} \! \lambda
\exp\left[-(\gamma_{k}\lambda)^{\alpha_{k}}\right]
J_{0}(r_{n}\lambda) \, d\lambda
\end{equation*}
where $\alpha_k$ and $\gamma_k$ are the parameters associated with the $k$th class and where $J_{0}$ is the zeroth order Bessel function of the first kind. Note that this distribution has been also used to model SAR images in \cite{Kuruoglu2004,Achim2006}. The prior distributions assigned to the parameters $\alpha_k$ and $\gamma_k$ ($k = 1,\ldots,K$) are uniform and inverse gamma distributions as in \cite{PereyraTMIC2011}. The estimation of $\beta$, $\boldsymbol{z}$ and $\boldsymbol{\theta} = (\boldsymbol{\alpha}^T,\boldsymbol{\gamma}^T)^T = (\alpha_1,\ldots,\alpha_K,\gamma_1,\ldots,\gamma_K)^T$ is performed by using Algorithm \ref{algo:hybridGibbs}. The sampling strategies described in Sections \ref{ssec:4.1} and \ref{subsec:auxiliary} can be used for the generation of samples according to $P[\boldsymbol{z}| \boldsymbol{\alpha},\boldsymbol{\gamma},\beta,\boldsymbol{r}]$ and $f(\beta|\boldsymbol{\alpha},\boldsymbol{\gamma},\boldsymbol{z},\boldsymbol{r})$. More details about simulation according to $f(\boldsymbol{\alpha}|\boldsymbol{\gamma},\boldsymbol{z},\beta, \boldsymbol{r})$ and $f(\boldsymbol{\gamma}|\boldsymbol{\alpha},\boldsymbol{z},\beta, \boldsymbol{r})$ are provided in the technical report \cite{Pereyra_TIP_TechReport_2012}..

\begin{figure}[htb]
\begin{minipage}[a1]{.49\linewidth}
  \centering
  \centerline{\includegraphics[width=4.0cm]{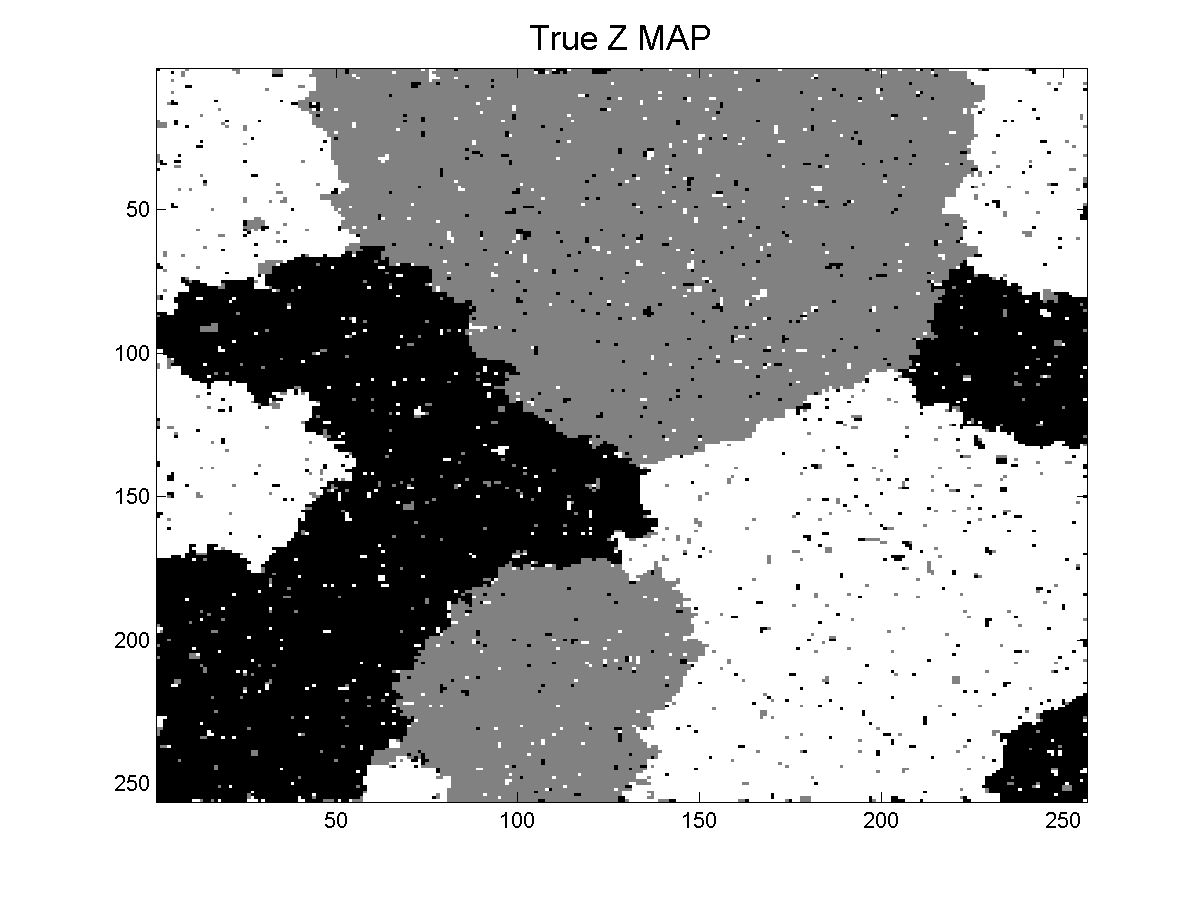}}
  \centerline{(a) \small{True Labels ($\beta = 1.2$)}}\medskip
\end{minipage}
\hfill
\begin{minipage}[a1]{.49\linewidth}
  \centering
  \centerline{\includegraphics[width=4.0cm]{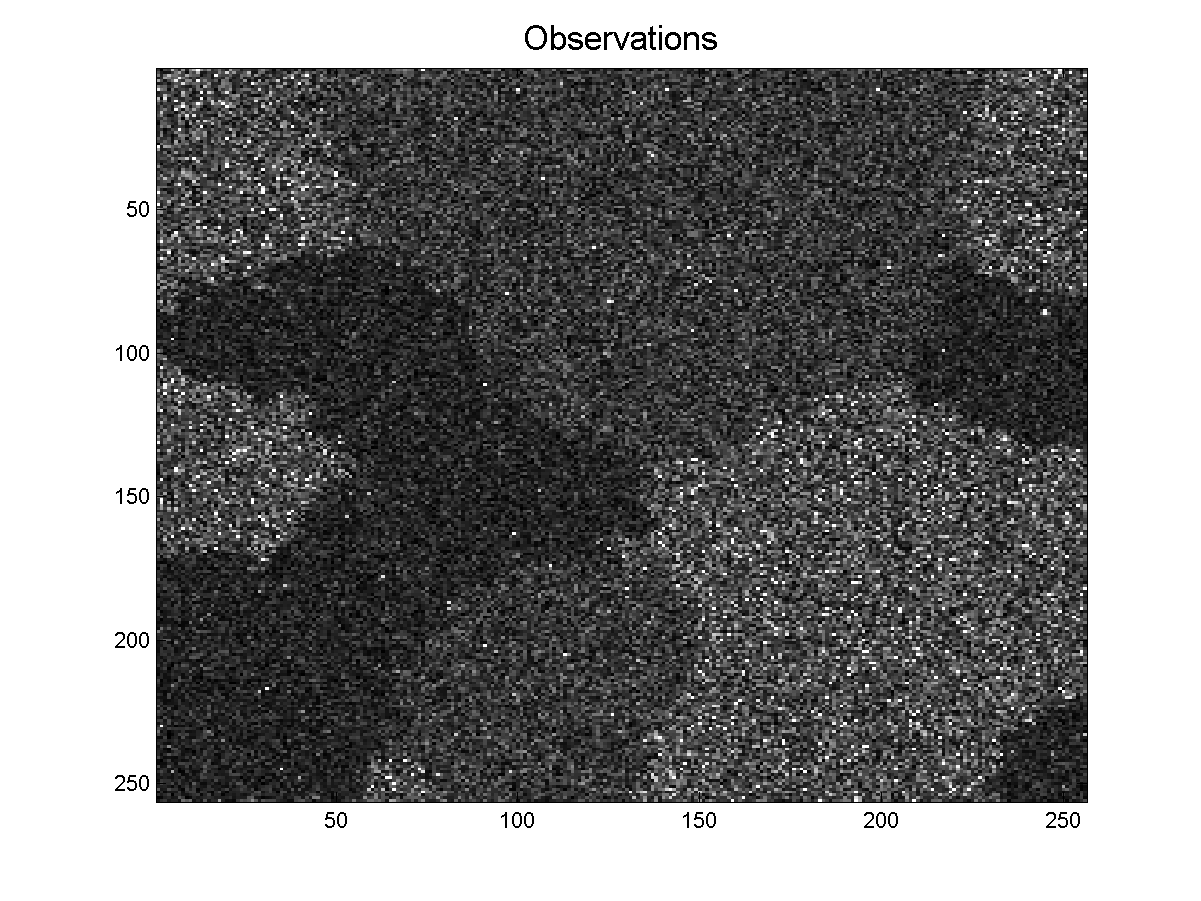}}
  \centerline{(b)  \small{Observations}}\medskip
\end{minipage}
\hfill
\begin{minipage}[a1]{.49\linewidth}
  \centering
  \centerline{\includegraphics[width=4.0cm]{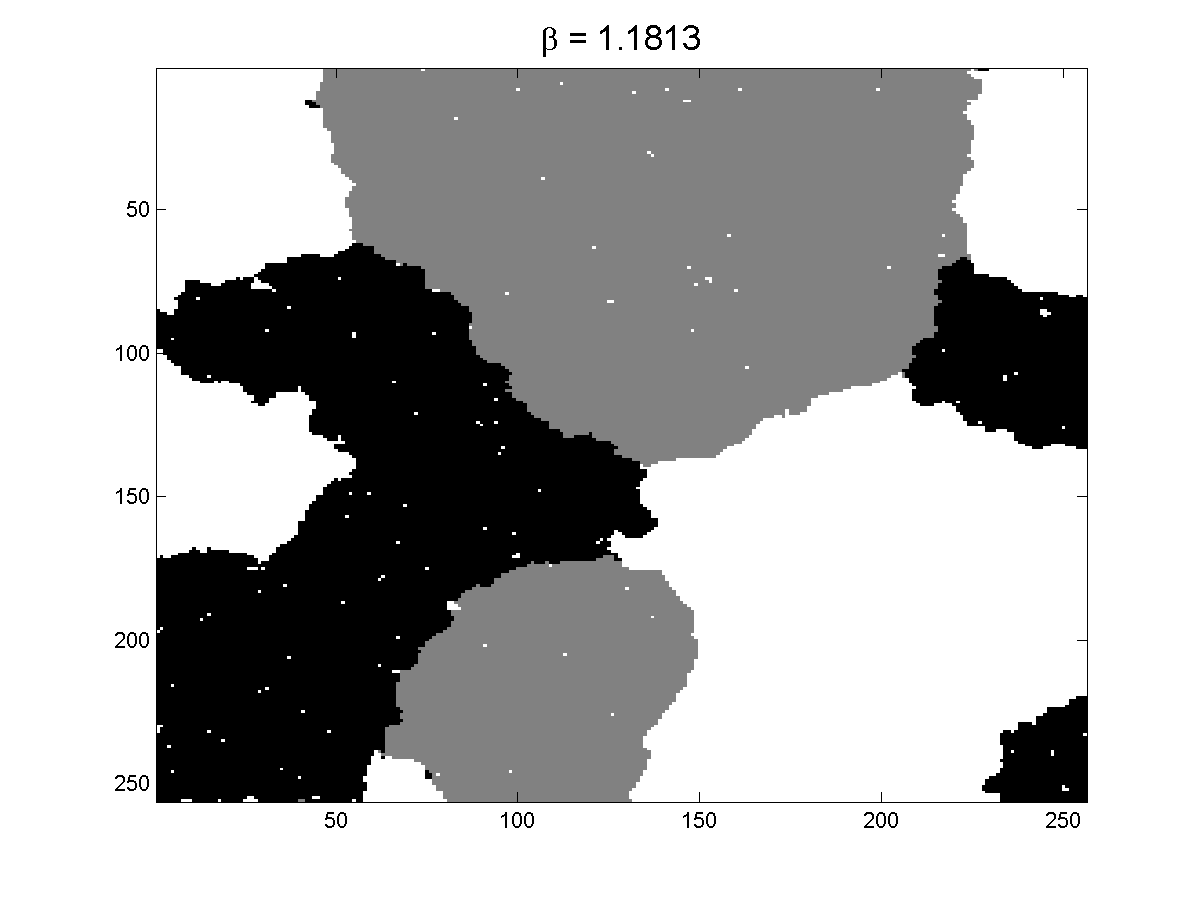}}
  \centerline{(c)  \small{Estimated $\beta$}}\medskip
\end{minipage}
\hfill
\begin{minipage}[a2]{.49\linewidth}
  \centering
  \centerline{\includegraphics[width=4.0cm]{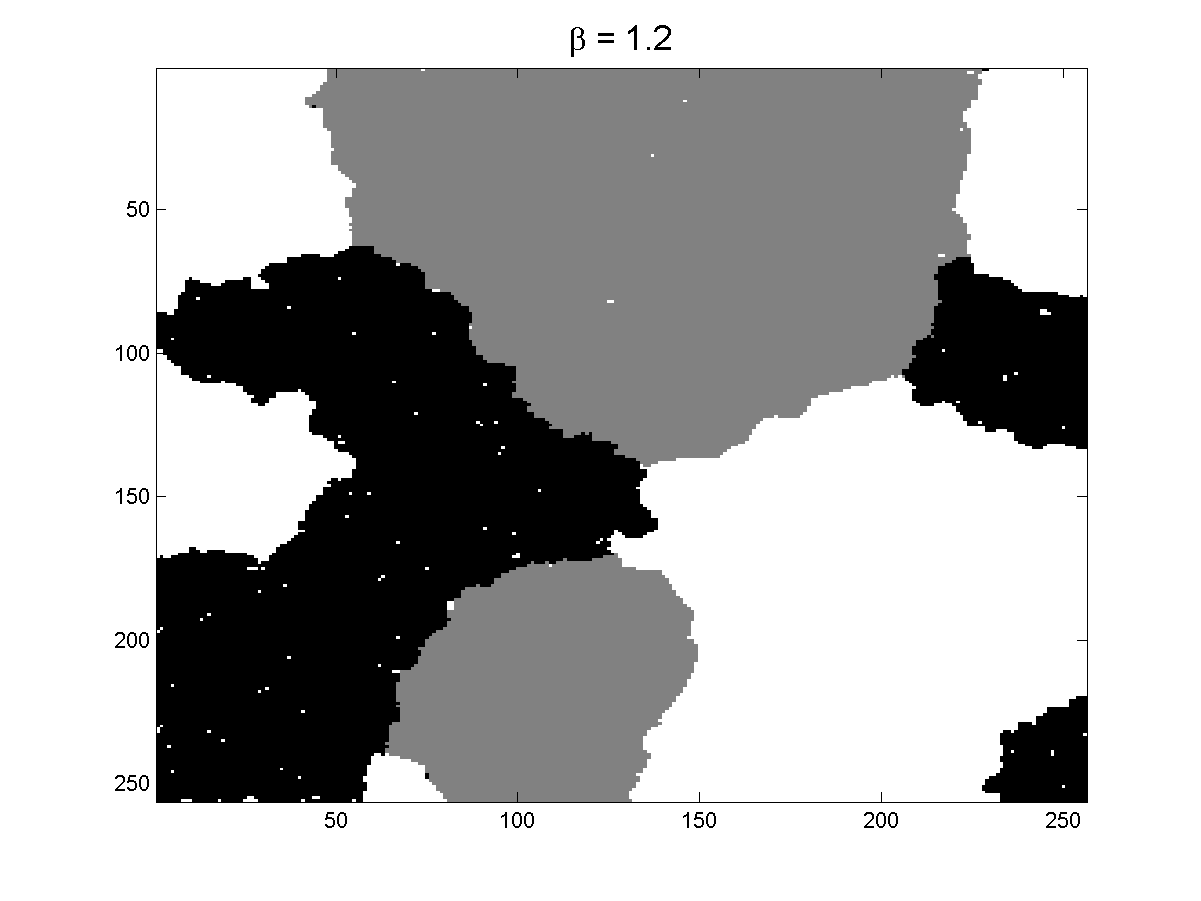}}
  \centerline{(d)  \small{True $\beta = 1.2$}}\medskip
\end{minipage}

\begin{minipage}[a2]{.49\linewidth}
  \centering
  \centerline{\includegraphics[width=4.0cm]{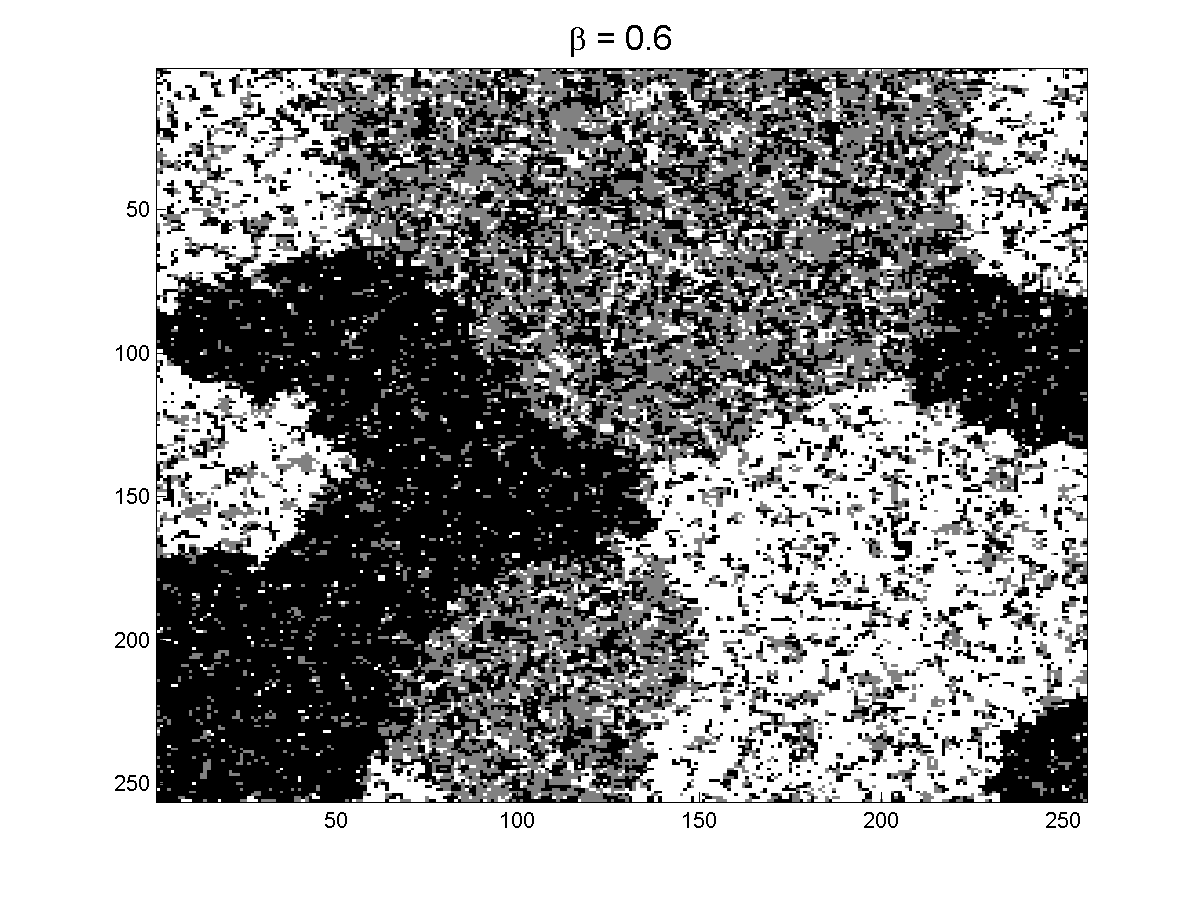}}
  \centerline{(e)  \small{$\beta = 0.6$}}\medskip
\end{minipage}
\hfill
\begin{minipage}[a2]{.49\linewidth}
  \centering
  \centerline{\includegraphics[width=4.0cm]{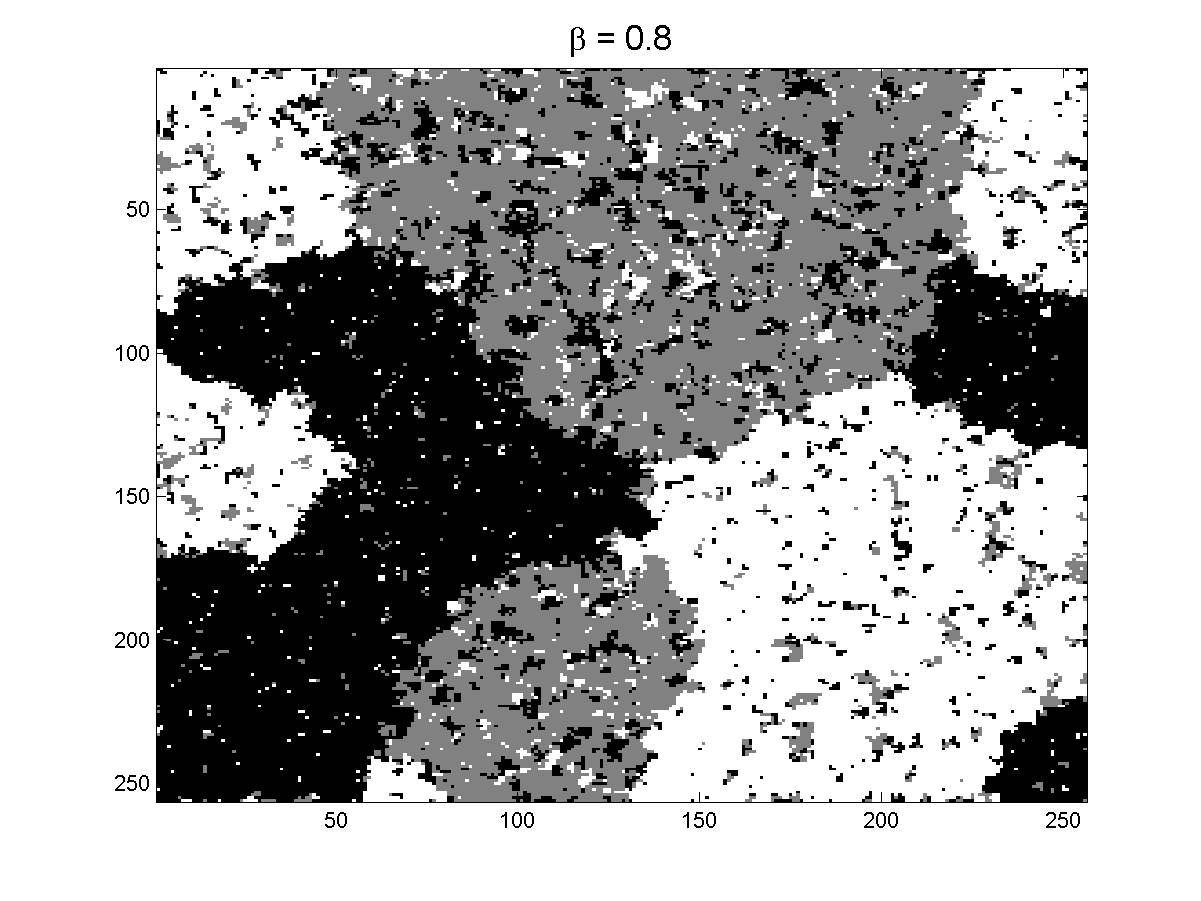}}
  \centerline{(f) \small{$\beta = 0.8$}}\medskip
\end{minipage}
\hfill
\begin{minipage}[a2]{.49\linewidth}
  \centering
  \centerline{\includegraphics[width=4.0cm]{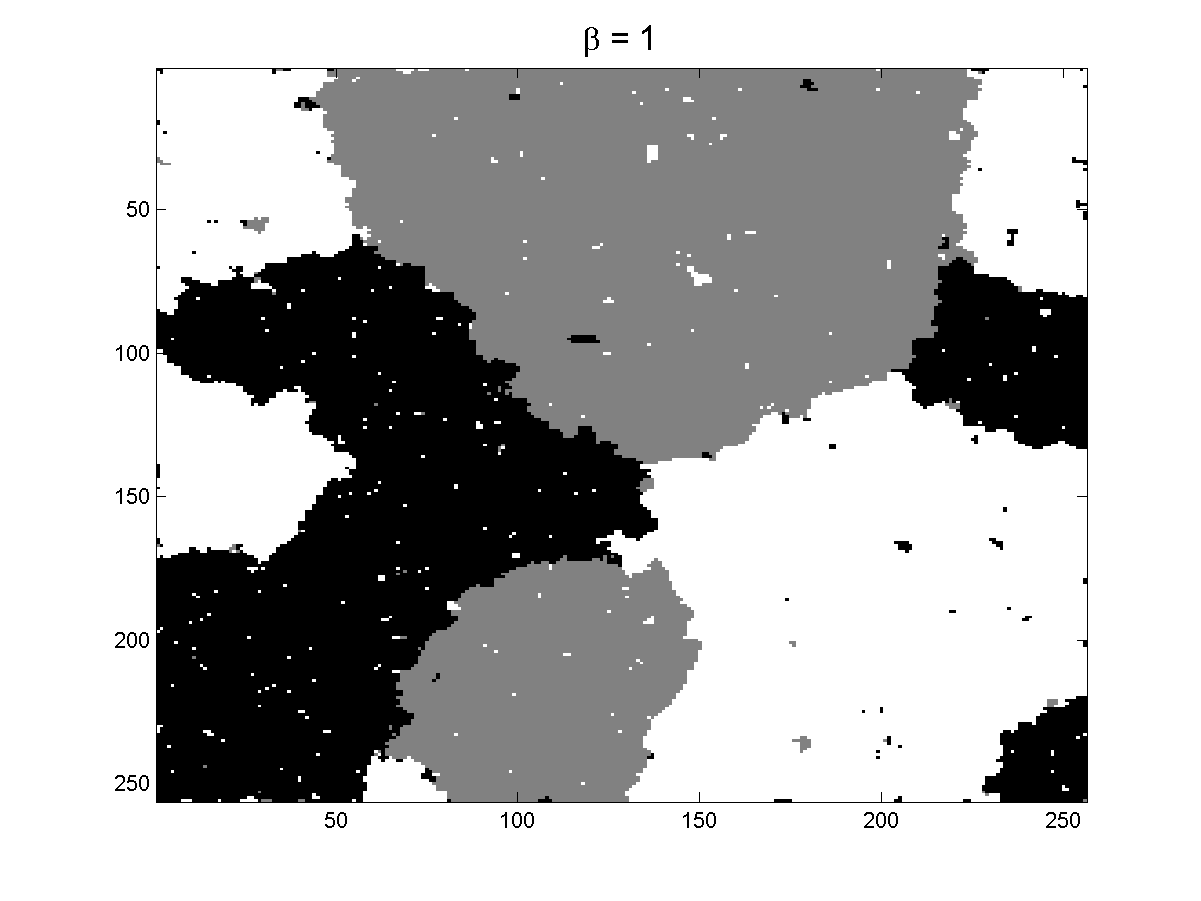}}
  \centerline{(g)  \small{$\beta = 1.0$}}\medskip
\end{minipage}
\hfill
\begin{minipage}[a2]{.49\linewidth}
  \centering
  \centerline{\includegraphics[width=4.0cm]{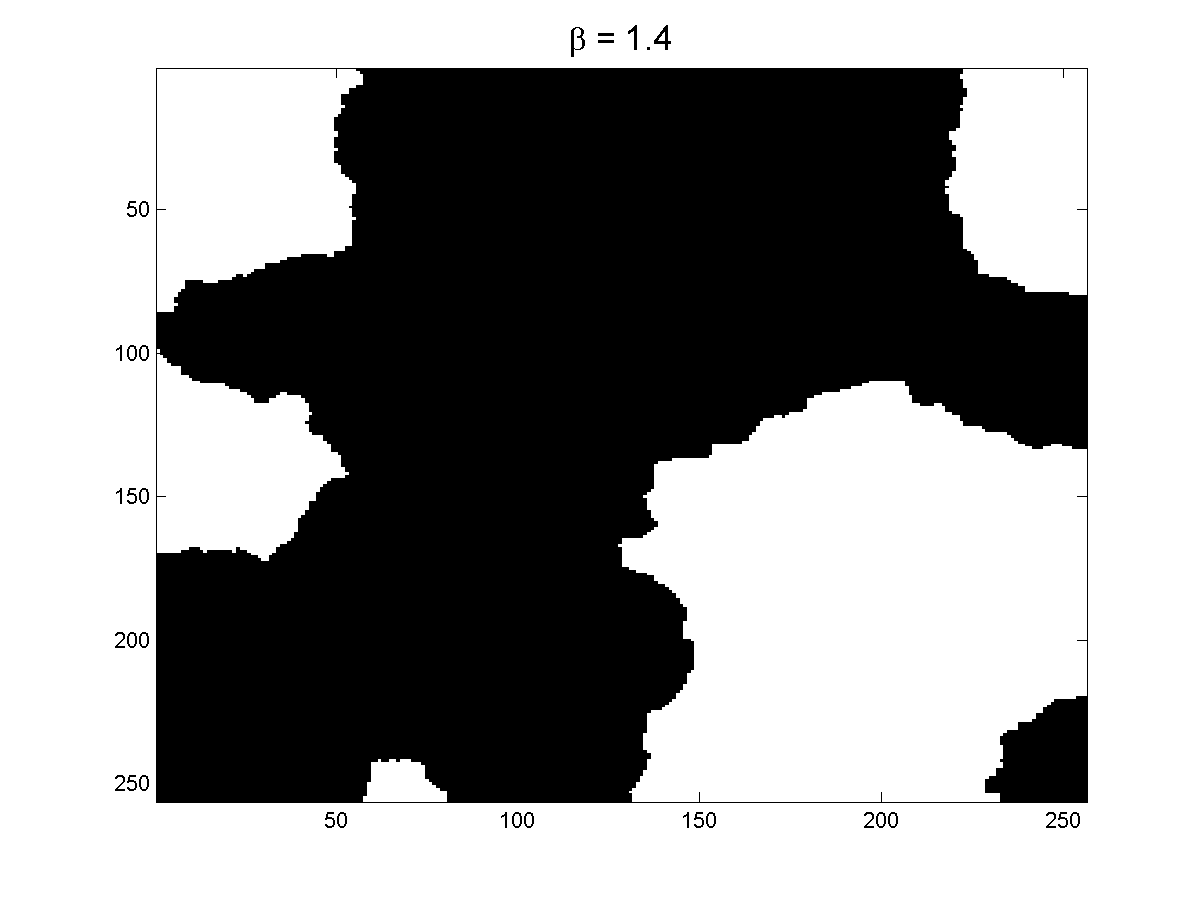}}
  \centerline{(h)  \small{$\beta = 1.4$}}\medskip
\end{minipage}
\caption{\small{$\alpha$-Rayleigh mixture: MAP estimates of the class labels. (a) Ground truth, (b) observations, (c) proposed algorithm (estimated $\beta$),(d) true $\beta = 1.2$, (e)-(h) fixed $\beta$ = (0.6, 0.8, 1.0, 1.2, 1.4).}}
\label{fig:betaPlot2}
\end{figure}

The following results have been obtained for a 3-component $\alpha$-Rayleigh mixture with parameters $\boldsymbol{\alpha} = (1.99 ; 1.99 ; 1.80)$ and $\boldsymbol{\gamma} = (1.0 ; 1.5 ; 2.0)$. Fig. \ref{fig:Gamma_aRayleigh_mixture}(b) shows the densities of the components associated with this $\alpha$-Rayleigh mixture. Again, note that there is significant overlap between the mixture components making the inference problem very challenging. For each experiment the MAP estimates of the class labels $\boldsymbol{z}$ have been computed from a single Markov chain of $T=2\,000$ iterations whose first $900$ iterations (burn-in period) have been removed. Table \ref{table4} shows the percentage of MAP class labels correctly estimated. The first column corresponds to labels that were estimated jointly with $\beta$ whereas the other columns result from fixing $\beta$ to different a priori values. To ease interpretation, the best and second best results for each simulation scenario in Table \ref{table4} are highlighted in red and blue. We observe that even if the mixture components are hard to estimate, the proposed method performs similarly to the case of a known coefficient $\beta$. Also, setting $\beta$ incorrectly degrades estimation performance considerably. Table \ref{table4b} shows the MMSE estimates of $\beta$, $\boldsymbol{\alpha}$ and $\boldsymbol{\gamma}$ corresponding to the three simulations of the first column of Table \ref{table4} (proposed method). We observe that these values are in good agreement with the true values used to generate the observation vectors. To conclude, Fig. \ref{fig:betaPlot2} shows the MAP estimates of the class labels corresponding to the simulation associated with the scenario reported in the last row of Table \ref{table4}. More precisely, the actual class labels are displayed in Fig. \ref{fig:betaPlot2}(a), which shows a realization of a 3-class Potts MRF with $\beta = 1.2$. The corresponding observation vector is presented in Fig. \ref{fig:betaPlot2}(b). Fig. \ref{fig:betaPlot2}(c) and Fig. \ref{fig:betaPlot2}(d) show the class labels obtained with the proposed method and with the actual value of $\beta$. Lastly, Figs. \ref{fig:betaPlot2}(e)-(h) show the results obtained when $\beta$ is fixed incorrectly to $0.6,0.8,1.0$ and $1.4$. We observe that the proposed method produces classification results that are very similar to those obtained when $\beta$ is fixed to its true value. On the other hand, fixing $\beta$ incorrectly generally leads to very poor results.

\begin{table}[htb]
\renewcommand{\arraystretch}{1.3}
\caption{$\alpha$-Rayleigh Mixture: Class label estimation ($K=3$)}
\label{table4}
\centering
\tabcolsep 5.0pt
\tiny
\begin{tabular}{|c||c|c|c|c|c|c|c|}
  \cline{4-8}
  \multicolumn{3}{c|}{} & \multicolumn{5}{c|}{Correct classification with $\beta$ fixed} \\
  \cline{2-8}
  \multicolumn{1}{c|}{} & \multicolumn{2}{c|}{Proposed method} & $\beta=0.6$ & $\beta=0.8$ & $\beta=1.0$ & $\beta=1.2$ & $\beta=1.4$ \\
  \hline
  \hline
  $\textrm{True } \beta = 0.8$ & $\hat{\beta}=0.82$ & \red56.5\% & 52.3\% & \blue56.3\% & 44.8\% & 33.3\% & 33.4\% \\
  \hline
  $\textrm{True } \beta = 1.0$ & $\hat{\beta}=1.01$ & \blue75.5\% & 61.1\% & 68.1\% & \red75.5\% & 54.1\% & 41.7\% \\
  \hline
  $\textrm{True } \beta = 1.2$ & $\hat{\beta}=1.18$ & \red95.0\% & 67.7\% & 83.1\% & 94.4\% & \blue94.8\% & 69.5\% \\
  \hline
\end{tabular}
\end{table}

\begin{table}[htb]
\renewcommand{\arraystretch}{1.3}
\caption{$\alpha$-Rayleigh Mixture: Parameter estimation}
\label{table4b}
\centering
\tabcolsep 5.0pt
\tiny
\begin{tabular}{|c||c|c||c|c||c|c|}
  \cline{2-7}
  \multicolumn{1}{c|}{} & true & MMSE & true & MMSE & true & MMSE \\
  \hline
  \hline
  $\beta$   & 0.80  & 0.81 $\pm$ 0.013   & 1.00 & 1.01 $\pm$ 0.015  & 1.20 & 1.18   $\pm$ 0.021\\
  \hline
  $\alpha_1$ & 1.99 & 1.98 $\pm$ 0.010   & 1.99 & 1.99 $\pm$ 0.010  & 1.99&  1.99   $\pm$ 0.004\\
  $\gamma_1$ & 1.00 & 1.00 $\pm$ 0.009   & 1.00 & 1.00 $\pm$ 0.009  & 1.00&  1.00   $\pm$ 0.005\\
  \hline
  $\alpha_2$ & 1.99 & 1.99 $\pm$ 0.007   & 1.99 & 1.97 $\pm$ 0.008  & 1.99&  1.99   $\pm$ 0.005\\
  $\gamma_2$ & 1.50 & 1.47 $\pm$ 0.012   & 1.50 & 1.49 $\pm$ 0.010  & 1.50&  1.50   $\pm$ 0.005\\
  \hline
  $\alpha_3$ & 1.80 & 1.80 $\pm$ 0.008   & 1.80 & 1.80 $\pm$ 0.006  & 1.80 & 1.79   $\pm$ 0.007\\
  $\gamma_3$ & 2.00 & 2.02 $\pm$ 0.014   & 2.00 & 1.97 $\pm$ 0.017  & 2.00 & 2.00   $\pm$ 0.009\\
  \hline
\end{tabular}
\end{table}

\section{Application to real data} \label{sec:Application to real data}
After validating the proposed Gibbs sampler on synthetic data, this section presents two applications of the proposed algorithm to real data.

\subsection{Pixel classification of a $2$D SAR image}\label{ssec:SAR}
\begin{figure}[htb]
\begin{minipage}[a1]{.49\linewidth}
  \centering
  \centerline{\includegraphics[width=4.0cm]{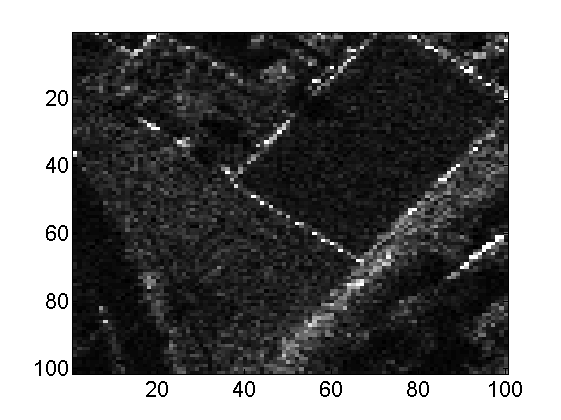}}
  \centerline{(a)  \small{Multilook SAR Image}}\medskip
\end{minipage}
\hfill
\begin{minipage}[a1]{.49\linewidth}
  \centering
  \centerline{\includegraphics[width=4.0cm]{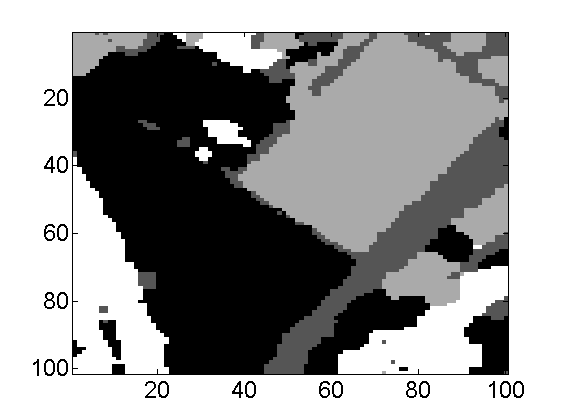}}
  \centerline{(b)  \small{Labels ($\hat{\beta} = 1.62$)}}\medskip
\end{minipage}

\begin{minipage}[a2]{.49\linewidth}
  \centering
  \centerline{\includegraphics[width=4.0cm]{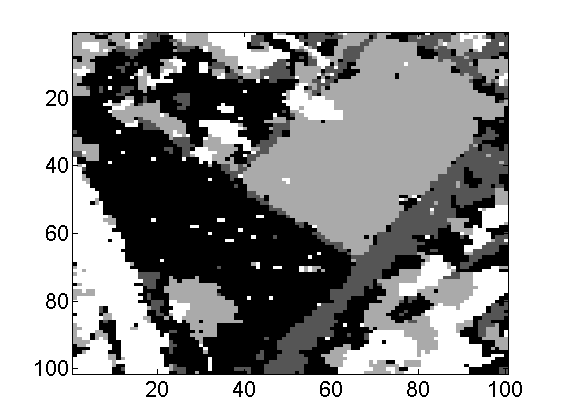}}
  \centerline{(c)  \small{Labels ($\beta$=1)}}\medskip
\end{minipage}
\hfill
\begin{minipage}[a2]{.49\linewidth}
  \centering
  \centerline{\includegraphics[width=4.0cm]{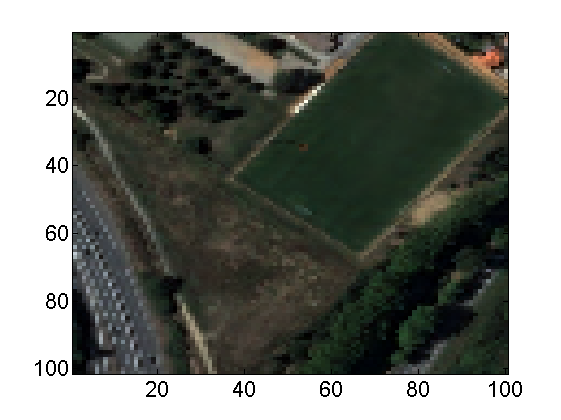}}
  \centerline{(d) \small{Optical Image of Toulouse}}\medskip
\end{minipage}

\caption{\small{Pixel classification in a multilook SAR image (c). MAP labels when $\beta$ is estimated (d) and $\beta = 1$ (e). Figs. (a)-(b) provide optical images of the same region.}}
\label{fig:SAR}
\end{figure}

The proposed method has been applied to the unsupervised classification of a $2$D multilook SAR image acquired over Toulouse, France, depicted in Fig. \ref{fig:SAR}(a). This image was acquired by the TerraSAR-X satellite at $1$m resolution and results from summing 3 independent SAR images (i.e., $L=3$). Potts MRFs have been extensively applied to SAR image segmentation using different observations models \cite{Tison2004,Deng2005,Yongfeng2005,Yu2010}. For simplicity the observation model chosen in this work is a mixture of gamma distributions (see Section \ref{ssec:mixture_gamma} and the report \cite{Pereyra_TIP_TechReport_2012} for more details about the gamma mixture model). The proposed experiments were conducted with a number of classes $K = 4$ (setting $K>4$ resulted in empty classes). Fig. \ref{fig:SAR}(b) shows the results obtained with the proposed method. The MMSE estimate of the granularity coefficient corresponding to this result is $\hat{\beta} = 1.62 \pm 0.05$, which has enforced the appropriate amount of spatial correlation to handle noise and outliers while preserving contours. Fig. \ref{fig:SAR}(c) shows the results obtained by fixing $\beta = 1$, as proposed in \cite{Deng2005}. These results have been computed from a single Markov chain of $T=5\,000$ iterations whose first $1\,000$ iterations (burn-in period) have been removed. Finally, for visual interpretation Fig. \ref{fig:SAR}(d) shows the same region observed by an airborne optical sensor. We observe that the classification obtained with the proposed method has clear boundaries and few miss-classifications.

\subsection{Lesion segmentation in a $3$D ultrasound image}\label{ssec:US}
The proposed method has also been applied to the segmentation of a skin lesion in a dermatological $3$D ultrasound image. Ultrasound-based lesion inspection is an active topic in dermatological oncology, where patient treatment depends mainly on the depth of the lesion and the number of skin layers it has invaded. This problem has been recently addressed using an $\alpha$-Rayleigh mixture model \eqref{eq:palphaR} coupled with a tridimensional Potts MRF as prior distribution for the class labels \cite{PereyraTMIC2011}. The algorithm investigated in \cite{PereyraTMIC2011} estimates the label vector and the mixture parameters conditionally to a known value of $\beta$ that is set heuristically by cross-validation. The proposed method completes this approach by including the estimation of $\beta$ into the segmentation problem. Some elements of this model are recalled in the technical report \cite{Pereyra_TIP_TechReport_2012}.

\begin{figure}[htb]
  \centering
  \centerline{\includegraphics[width=5.0cm]{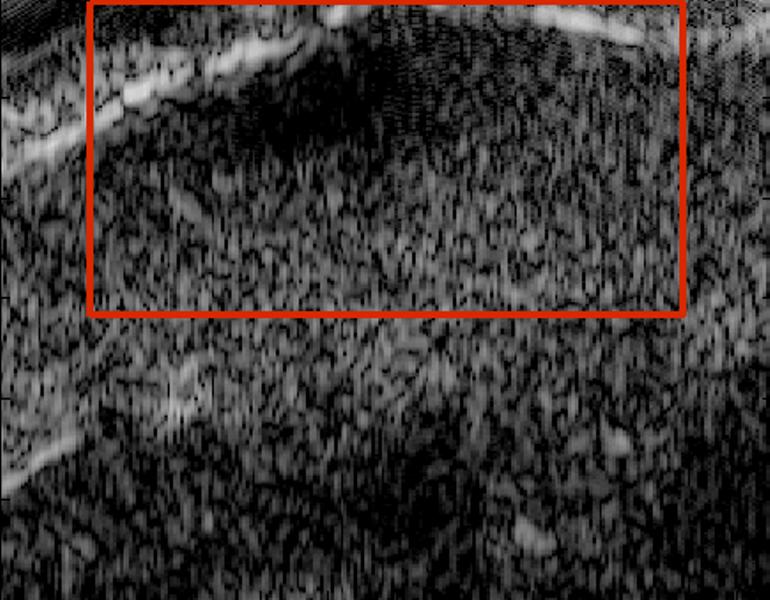}}
  \centerline{(a) \small{Dermis view with skin lesion (ROI $= 160 \times 175 \times 16$).}}\medskip
\begin{minipage}[a2]{.49\linewidth}
  \centering
  \centerline{\includegraphics[width=4.0cm]{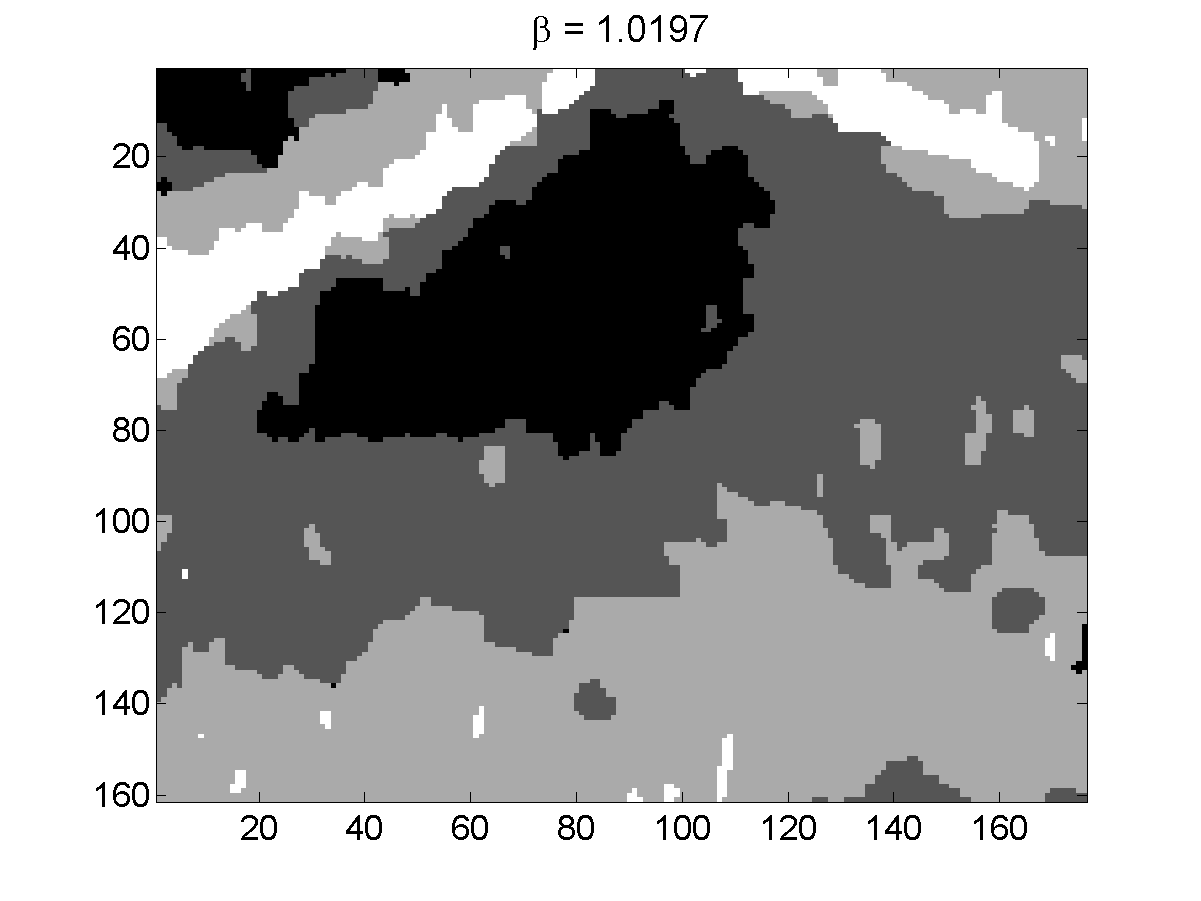}}
  \centerline{(b)  \small{(Estimated $\hat{\beta}$)}}\medskip
\end{minipage}
\hfill
\begin{minipage}[a2]{.49\linewidth}
  \centering
  \centerline{\includegraphics[width=4.0cm]{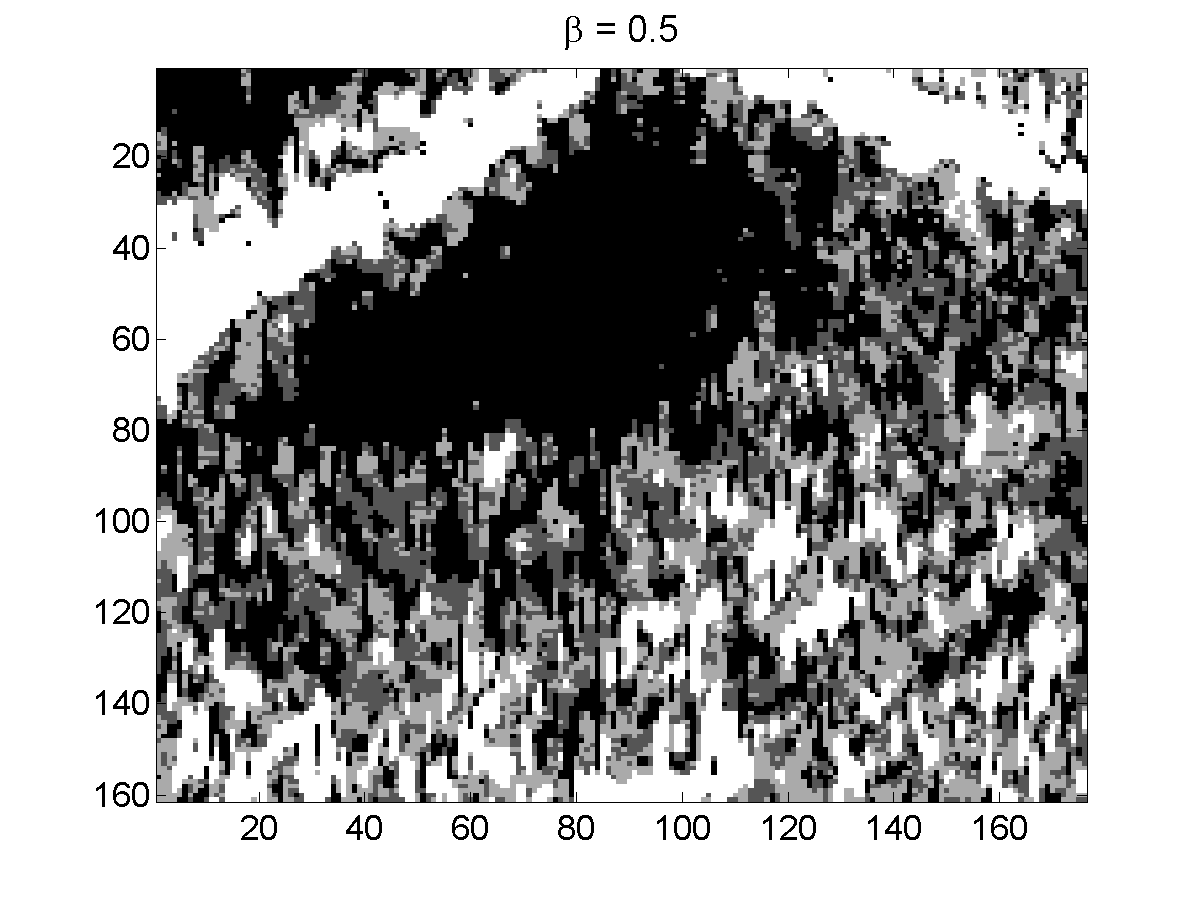}}
  \centerline{(c)  \small{($\beta = 0.5$)}}\medskip
\end{minipage}

\begin{minipage}[a2]{.49\linewidth}
  \centering
  \centerline{\includegraphics[width=4.0cm]{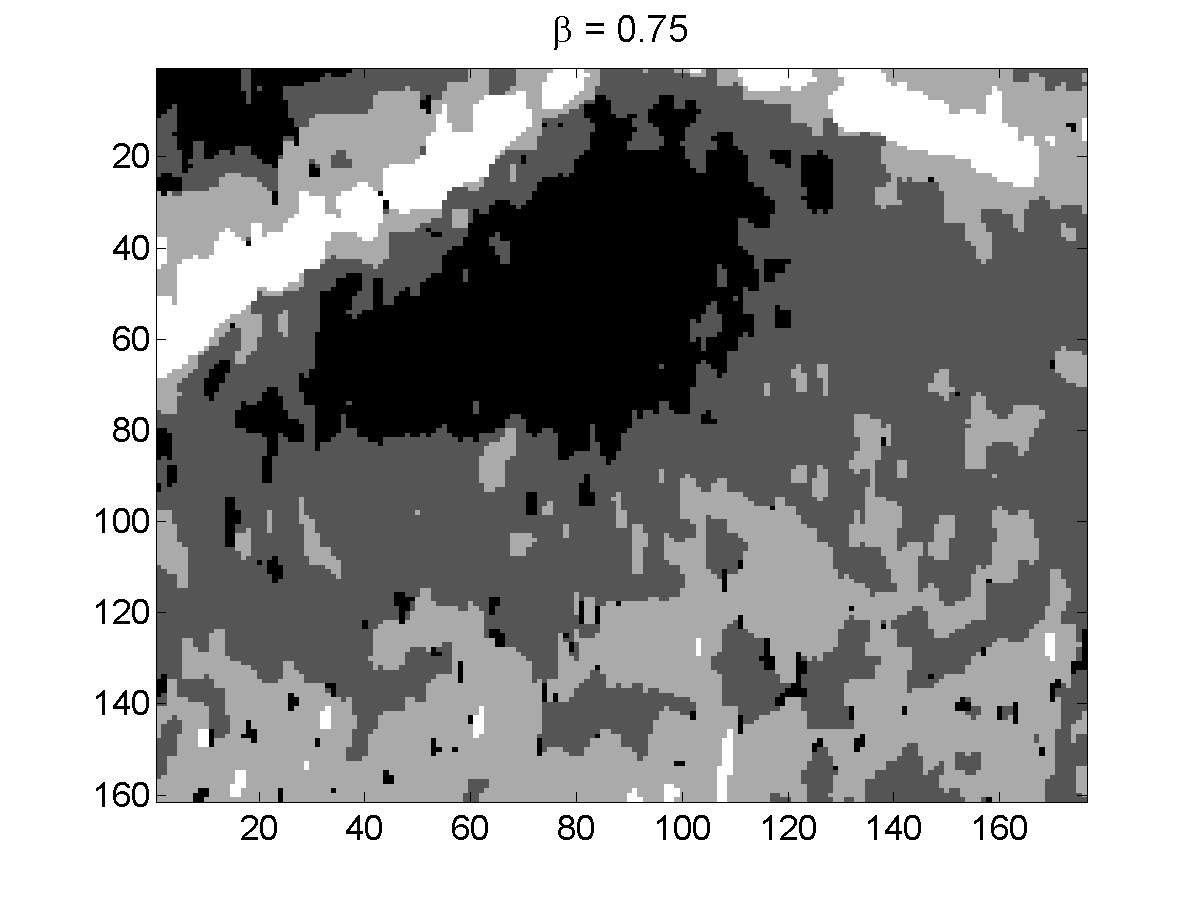}}
  \centerline{(d) \small{($\beta = 0.75$)}}\medskip
\end{minipage}
\hfill
\begin{minipage}[a1]{.49\linewidth}
  \centering
  \centerline{\includegraphics[width=4.0cm]{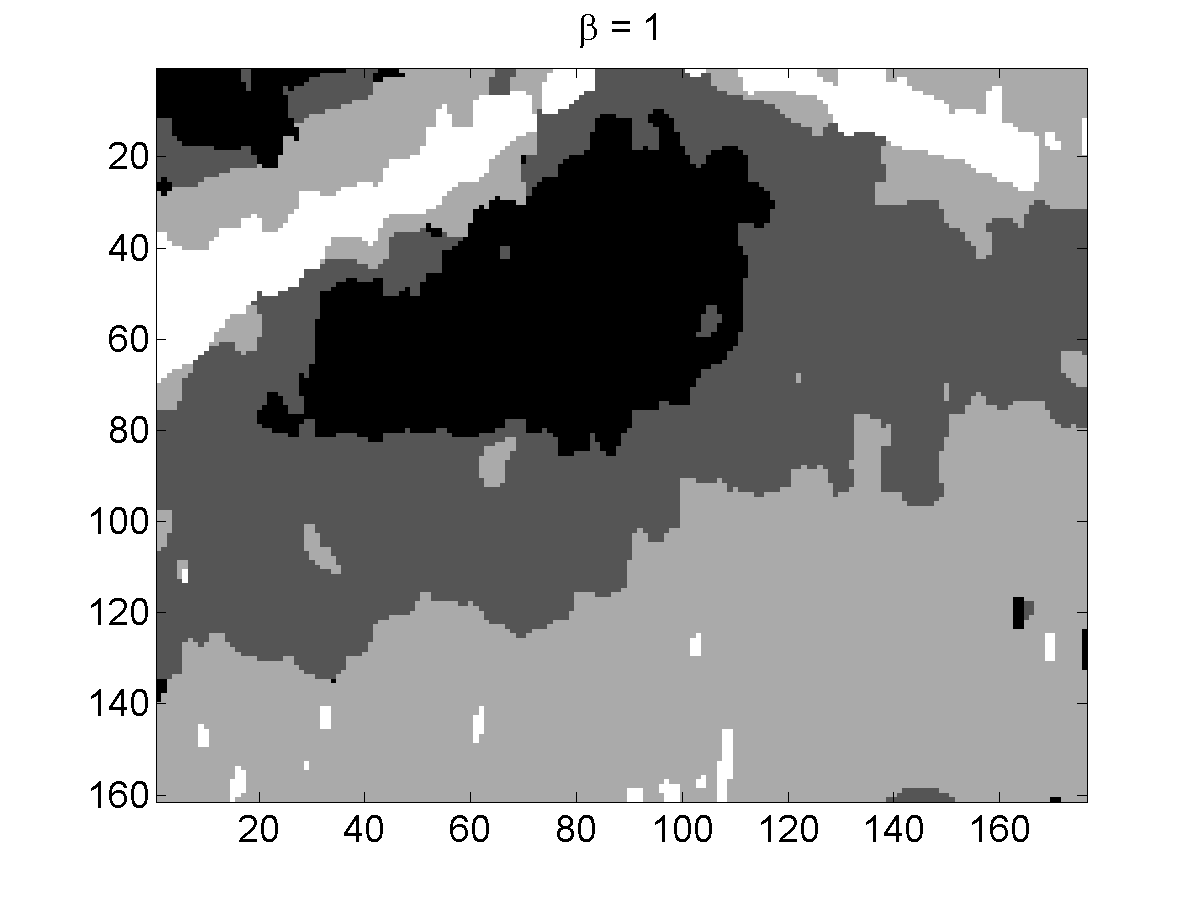}}
  \centerline{(e)  \small{($\beta = 1.0$)}}\medskip
\end{minipage}

\begin{minipage}[a2]{.49\linewidth}
  \centering
  \centerline{\includegraphics[width=4.0cm]{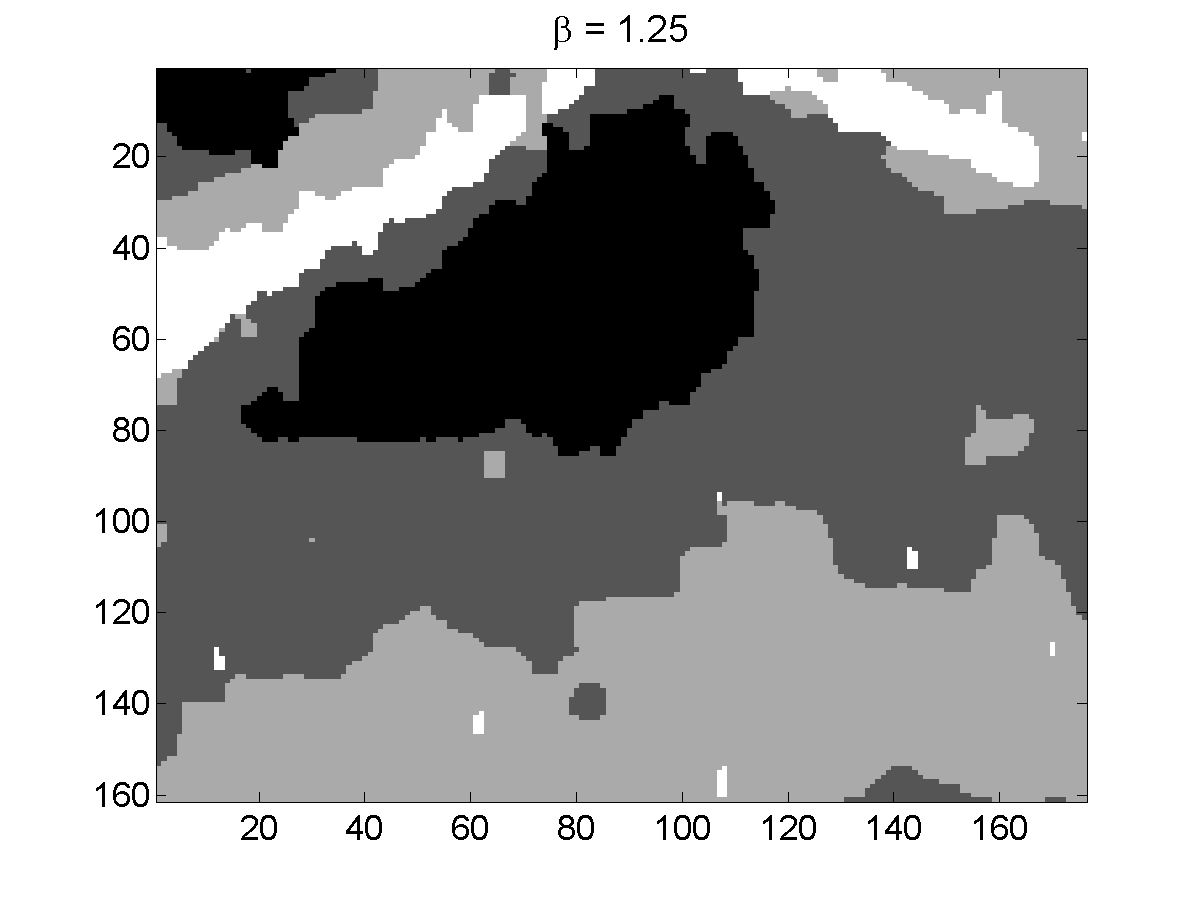}}
  \centerline{(f)  \small{($\beta = 1.25$)}}\medskip
\end{minipage}
\hfill
\begin{minipage}[a2]{.49\linewidth}
  \centering
  \centerline{\includegraphics[width=4.0cm]{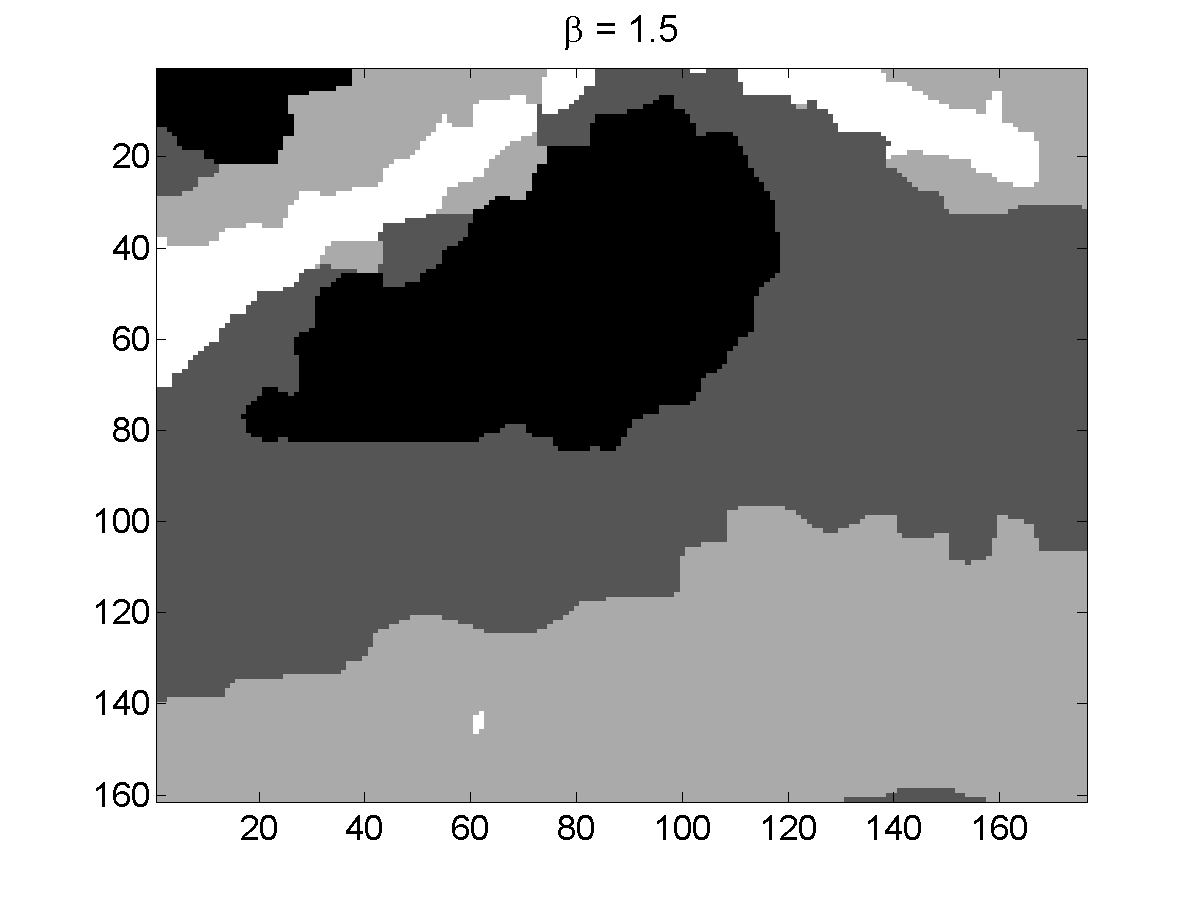}}
  \centerline{(g)  \small{($\beta = 1.5$)}}\medskip
\end{minipage}
\caption{\small{Log-compressed US images of skin lesion and the corresponding estimated class labels (\textit{lesion} = black,
\textit{epidermis} = white, \textit{pap. dermis} = dark gray, \textit{ret. dermis} = light gray). MAP estimates of the class labels. Fig. (b) shows the results obtained r
with the proposed method.  Figs. (c)-(g) show the results obtained with the algorithm \cite{PereyraTMIC2011} for $\beta = (0.5,0.75,1,1.25,1.5)$.}}
\label{fig:USImage}
\end{figure}

Fig. \ref{fig:USImage}(a) shows a $3$D B-mode ultrasound image of a skin lesion, acquired at $100$MHz with a focalized $25$MHz $3$D probe (the lesion is contained within the ROI outlined by the red rectangle). Fig. \ref{fig:USImage}(b) presents one slice of the $3$D MAP label vector obtained with the proposed method. The MMSE estimate of the granularity coefficient corresponding to this result is $\hat{\beta} = 1.020 \pm 0.007$. To assess the influence of $\beta$, Figs. \ref{fig:USImage}(c)-(g) show the MAP class labels obtained with the algorithm proposed in \cite{PereyraTMIC2011} for different values of $\beta$. These results have been computed using $K = 4$ since the region of interest (ROI) contains 3 types of healthy skin layers (epidermis, papillary dermis and reticular dermis) in addition to the lesion. Labels have been computed from a single Markov chain of $T=12\,000$ iterations whose first $2\,000$ iterations (burn-in period) have been removed.

\begin{figure}[htb]
\centering
\centerline{\includegraphics[width=4.5cm]{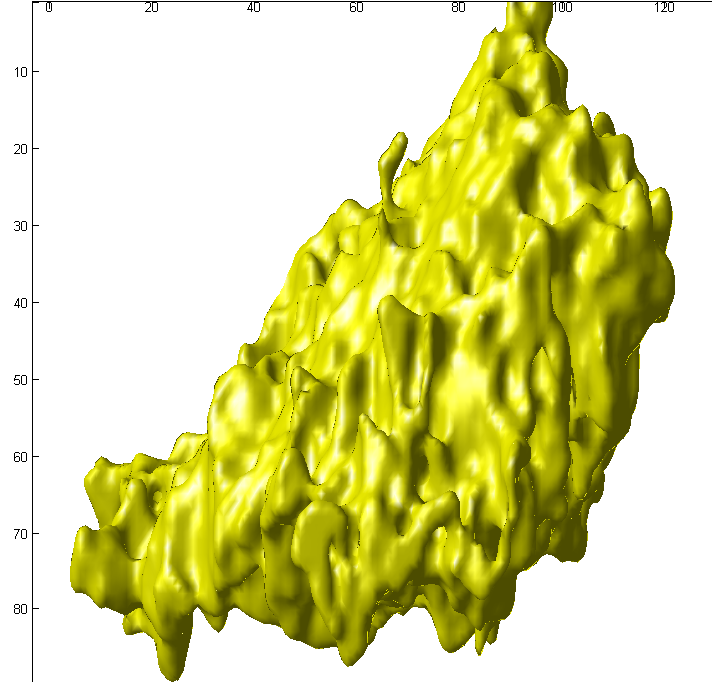}}
\caption{\small{Frontal viewpoint of a $3$D reconstruction of the skin lesion.}}
\label{fig:3D_USImage}
\end{figure}

We observe that the proposed method produces a very clear segmentation that not only sharply locates the lesion but also provides realistic boundaries for the healthy skin layers within the ROI. This result indicates that the lesion, which is known to have originated at the dermis-epidermis junction, has already invaded the upper half of the papillary dermis. We also observe that the results obtained by fixing $\beta$ to a small value are corrupted by ultrasound speckle noise and fail to capture the different skin layers. On the other hand, choosing a too large value of $\beta$ enforces excessive spatial correlation and yields a segmentation with artificially smooth boundaries. Finally, Fig. \ref{fig:3D_USImage} shows a frontal viewpoint of a $3$D reconstruction of the lesion surface. We observe that the tumor has a semi-ellipsoidal shape which is cut at the upper left by the epidermis-dermis junction. The tumor grows from this junction towards the deeper dermis, which is at the lower right.

\section{Concluding Remarks}
\label{sec:conclusion} This paper presented a hybrid Gibbs sampler for estimating the Potts parameter $\beta$ jointly with the unknown parameters of a Bayesian model. In most image processing applications this important parameter is set heuristically by cross-validation. Standard MCMC methods cannot be applied to this problem because performing inference on $\beta$ requires computing the intractable normalizing constant of the Potts model. In this work the estimation of $\beta$ has been included within an MCMC method using an ABC likelihood-free Metropolis-Hastings algorithm, in which intractable terms have been replaced by simulation-rejection schemes. The ABC distance function has been defined using the Potts potential, which is the natural sufficient statistic of the Potts model. The proposed method can be applied to large images both in $2$D and in $3$D scenarios. Experimental results obtained for synthetic data showed that estimating $\beta$ jointly with the other unknown parameters leads to estimation results that are as good as those obtained with the actual value of $\beta$. On the other hand, choosing an incorrect value of $\beta$ can degrade the estimation performance significantly. Finally, the proposed algorithm was successfully applied to real bidimensional SAR and tridimensional ultrasound images. This study assumed that the number of classes $K$ is known. Future works could relax this assumption by studying the estimation of $\beta$ within a reversible jump MCMC algorithm \cite{Green1997} or by considering model choice ABC methods \cite{Grelaud2009}. Other perspectives for future work include the estimation of the total variation regularization parameter in image restoration problems \cite{Rudin1992} and the estimation of texture descriptors defined through Markov fields \cite{LiBook}.

\section*{Acknowledgment}
This work was developed as part of the CAMM4D project, funded by the French FUI and the Midi-Pyrenees region. We would like to acknowledge the support of the CNES, who provided the SAR and optical images used in Section \ref{ssec:SAR}. We are also grateful to the Hospital of Toulouse and Pierre Fabre Laboratories for the corpus of US images used in Section \ref{ssec:US}.

\bibliographystyle{ieeetran}
\bibliography{strings}

\end{document}